\documentclass[%
 reprint,   
superscriptaddress,   
 amsmath,   amssymb,   
 aps,   
prb,   
]{revtex4-2}

\newcommand{\dq}[1]{``#1''} 
\usepackage{graphicx}
\usepackage{dcolumn}
\usepackage{bm}
\usepackage{graphicx}  
\usepackage{subcaption}  
\usepackage{caption}  
\usepackage{braket} 
\usepackage[english]{babel}
\usepackage{float}

\begin{document}

\preprint{APS/123-QED}

\title{Two-electron edge states in a double SSH-chain of quantum dots }

\author{Keyu Wang}
\affiliation{Shenzhen Institute for Quantum Science and Engineering, Southern University of Science and Technology, Shenzhen 518055, China}
\affiliation{International Quantum Academy, Shenzhen 518048, China}
\affiliation{Guangdong Provincial Key Laboratory of Quantum Science and Engineering, Southern University of Science and Technology, Shenzhen 518055, China}
\author{Peihao Huang}%
 \email{huangpeihao@iqasz.cn}
\affiliation{Shenzhen Institute for Quantum Science and Engineering, Southern University of Science and Technology, Shenzhen 518055, China}
\affiliation{International Quantum Academy, Shenzhen 518048, China}
\affiliation{Guangdong Provincial Key Laboratory of Quantum Science and Engineering, Southern University of Science and Technology, Shenzhen 518055, China}

\date{\today}

\begin{abstract}
	
We study an interacting two-body model with adjustable  spin tunneling in the context of the double SSH chains for a quantum dot system. We discovered that varying interaction strengths and spin tunneling significantly influence the properties of correlated edge states in the energy spectrum obtained through exact diagonalization. We observe that stronger interactions lead to longer decay lengths of these states. Conversely, the decay length decreases as the difference in intracell or intercell tunneling increases. Importantly, the decay length is strongly correlated with the dynamical behavior of two particles; specifically, an increase in decay length corresponds to a decrease in motion frequency. This conclusion is supported by the observation of the expectation value of coordinate operators of the particles.

\end{abstract}

\maketitle

\section{Introduction}
Quantum simulators,    which utilize controllable quantum systems to model target systems,    have emerged as a vital tool for investigating many-body problems\cite{feynman_simulating_1982,   buluta_quantum_2009}.  The swift progress in quantum technology over the past two decades has empowered researchers to develop various artificial and controllable quantum platforms for probing new quantum states of matter and conducting quantum simulations. These platforms include superconducting circuits\cite{cui_ab_2025},    ultra-cold atomic gases\cite{bloch_quantum_2012,   he_recent_2025,   zhang_observation_2025},    ion traps\cite{zhang_observation_2017},    Rydberg atoms\cite{bernien_probing_2017},    photonic systems\cite{chen_research_2022},    and semiconductor quantum dots\cite{hensgens_quantum_2017}. Among these,    semiconductor quantum dot platforms have attracted considerable interest due to their exceptional tunability and scalability. 

Quantum dots are quasi-zero-dimensional semiconductor nanostructures capable of trapping electrons or holes. Their size,    dimension,    energy level structure,    and the number of trapped electrons can be controlled. Owing to the intrinsic long-range Coulomb interaction,    as well as the controllable tunneling coupling and lattice filling in quantum dot arrays,    this system can function as a microscale laboratory for investigating many-body physics\cite{kiczynski_engineering_2022}.  

A variety of quantum simulation studies on strongly correlated models have been performed using semiconductor quantum dot systems. These models include the Su-Schrieffer-Heeger(SSH) model\cite{kiczynski_engineering_2022}, Nagaoka ferromagnetism\cite{dehollain_nagaoka_2020},    antiferromagnetic Heisenberg Chain\cite{van_diepen_quantum_2021},    the Fermi-Hubbard model\cite{hensgens_quantum_2017,   byrnes_quantum_2008,   wang_experimental_2022},    and the excitonic two-channel Hubbard model\cite{PhysRevX.14.011048}. 

The investigation of two-body problems in quantum dot quantum simulation remains limited, although numerous studies have been conducted in other fields. People have explored the two-particle problem in the SSH model\cite{di_liberto_two-body_2016} and extended SSH Model\cite{di_liberto_two-body_2017,   marques_topological_2018},    as well as in one-dimensional optical lattices involving particles with distinct spins\cite{lin_interaction-induced_2020}. These studies demonstrate interesting topological properties

To date,    no theoretical simulations have been developed for two-body fermions in semiconductor quantum dot systems that consider spin-dependent tunneling,  the interaction steming from aligning  the same site simultaneously.  Due to the fact that the laboratory has already created the quantum dot ladder, which allows us to consider two-channel models. Therefore, the theoretical model proposed in this study is a two chains SSH Hubbard model for two-body fermions, which integrates spin-dependent tunneling and the interaction steming from aligning the same site.

By modulating the spin-dependent intercell and intracell tunneling strengths of electron,    transitions between different topological phases can be realized.  In the nontrivial topological phases,  the number and decay length of edge states affect the motion of electrons along the chain.  In addition, the interaction between electrons not only correlate their dynamics but also influence the topological properties of states, including the emergence of edge bound states. Therefore,    considering the combined effects of spin-dependent tunnelings and interactions,    we can find different topological phases through the decay length of the correlated edge states and  electron dynamics.  This work holds significant importance for experimentally probing electron dynamics and topological properties in a double array quantum dots and important implication for  many body physics in quantum dot system.

Our study focus on the correlated edge states observed in the energy spectrum of two-body systems, as well as the influence of interactions and tunneling strengths on their decay length and dynamics. Specifically, with other parameters held constant, a weaker interaction or a greater difference in tunneling strengths for different spins
results in a shorter decay length for the edge states of correlated two particle states and a reduced or more localized frequency of particle motion at the chain's edge. This conclusion was corroborated by plotting the oscillatory trajectories of two particles over time.

\subsection{Model}

\begin{figure}
	\centering
	\includegraphics[scale=0.49]{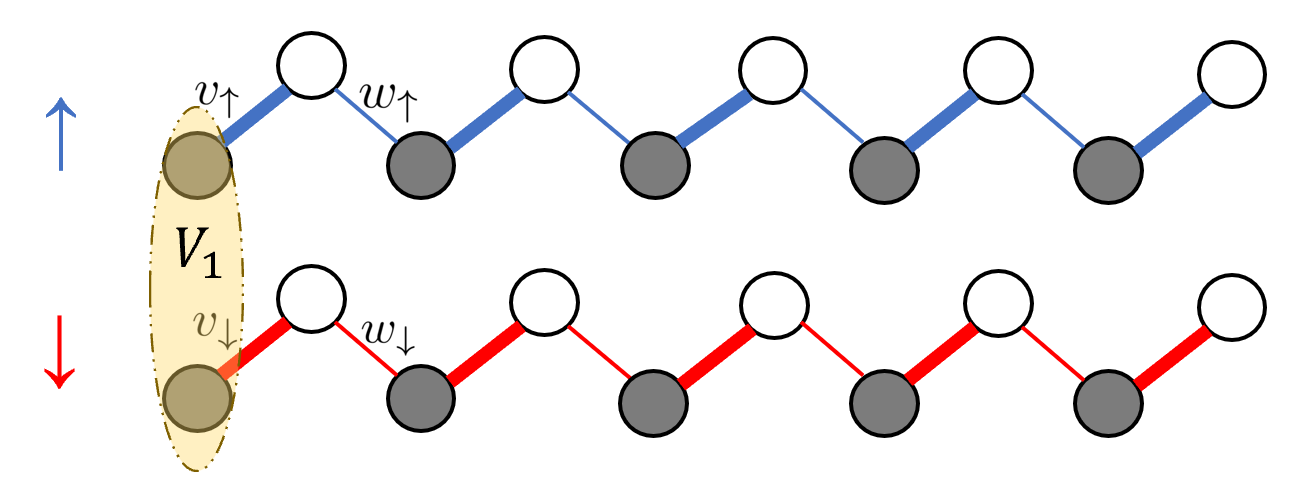}
	\captionsetup{justification=raggedright,    singlelinecheck=false}
	\caption{\label{3-1} Diagram of two arrays of quantum dots, where each array is an SSH chain. A spin-up electron is moving in upper array, while another spin-down electron is moving in the other array. The intracell tunnelings for the two electrons with different spins are denoted by $v_\uparrow$ and $v_\downarrow$ ,    while the intercell tunnelings are denoted by $w_\uparrow$ and $w_\downarrow$.  When the two electrons align at the same lattice site,    the Coulomb repulsion energy is $V_1$.  }
\end{figure}

In this work,    we consider a $2\times N$ atomic-doped quantum dots ladder to simulate a two-channel SSH Hubbard model,    where different channels are occupied by electrons with different spin. Electrons are only allowed to move on their respective chains, and cannot tunnel between two chains. Due to the close distance between the two chains, when two electrons align with the same site, it can be approximately equivalent to two electrons being at the same site. At this point, the Coulomb repulsion potential energy between the electrons is $V_1$, analogous to the Hubbard interaction, as shown in Fig. \ref{3-1}

The Hamiltonian of this two-body system can be written as 
\begin{equation}
	\begin{aligned}
		{H}=&  \sum_{l,   \sigma} v_{\sigma}(c_{l \sigma,   A}^{\dagger} c_{l \sigma,   B}+\mathrm{H. c. } )+ \sum_{l,   \sigma}w_{\sigma}  (c_{l \sigma,   B}^{\dagger} c_{l+1 \sigma,   A}+\mathrm{H. c. } ) \\
		& +V_1 \sum_{l } (n_{l\uparrow,   A } n_{l\downarrow,   A} +n_{l\uparrow,   B } n_{l\downarrow,   B}) ).
	\end{aligned}
\end{equation}

Here,    $c_{l \sigma,   A(B)}^{\dagger}$,   $c_{l \sigma,   A(B)}$ are the creation and annihilation operators of fermions with spin $\sigma=\uparrow, \downarrow$ on the $A$ or $B$ sublattice at the $l$-th unit cell.  $n_{l\sigma,   A(B) } =c_{l \sigma,   A(B)}^{\dagger}c_{l \sigma,   A(B)}$ is the number operator for a particle.  $v_{\sigma}$ and $w_{\sigma}$ describe electron intracell hopping and intercell hopping strength respectively.  $V_1$ is the effective on-site Coulomb repulsion energy. Assuming there are $N$ unit cells,    the total chain length is $L=2N$. 

An important method for studying the two body problem is that the one-dimensional two-body problem can be transformed into a two-dimensional single-particle problem.  In real space,    the two-particle state can be described by the unit cells coordinate,    denoted as $\ket{r_{\uparrow},   Z_1; r_{\downarrow},   Z_2}$.  Here,    $r_{\uparrow}$ and $r_{\downarrow}$ represent the unit cell positions of the two particles,    and $Z_1(Z_2)$ represent the positions of sublattices.  Considering the periodic boundary conditions (PBC),    the unit cells have translational invariance,    allowing us to apply Bloch's theorem to the system.  The electron states in real space can transformed to momentum space through Fourier transformation,   
\begin{align}
	\left|r_{\uparrow \downarrow},   \left\{(Z_1,   Z_2)\right\}\right\rangle &= \frac{1}{\sqrt{L}} \sum_{l=1}^{N} e^{iKl}\left|r_{\uparrow}+l,   Z_1; r_{\downarrow}+l,   Z_2\right\rangle, \notag \\
	\left\{(Z_1,   Z_2)\right\} &= \left\{(A,   A); (A,   B); (B,   A); (B,   B)\right\}, \label{3. 1. 2}
\end{align} where the positions of the two electrons are described by their relative unit cell position $r_{\uparrow\downarrow}=r_{\downarrow}-r_{\uparrow}$ and $K$ represents the center of mass momentum of the two electrons.

To simplify the calculations,    we choose the gauge  where the electron with spin up is assumed to be at the first unit cell  $r_{\uparrow}=0$,    and the position of the electron with spin down is represented by the relative unit cell position $r_\downarrow=r_{\uparrow \downarrow}$, $r_{\downarrow\uparrow}\in[-L/2,   L/2]$. Then,    we expand the  Hamiltonian using the momentum space basis vectors $\ket{r_{\uparrow\downarrow }, {(Z_1,   Z_2)}}$,    resulting in a $4N$-dimensional Hamiltonian matrix. We arrange the basis in the order by $r_{\uparrow \downarrow}=-1, 0, 1, ..., L/2, ..., -L/2, -L/2$+$1,...-2$ and $(Z_1,   Z_2) = (A,   A), (A,   B), (B,   A), (B,   B)$. 

 The matrix form of the Hamiltonian is as follows:
\begin{equation} 
\begin{pmatrix}
	\mathcal{A}  &\mathcal{B} & 0&0 &...&\mathcal{B} \\
	\mathcal{B^\dagger}  & \mathcal{A'} & \mathcal{B}&0&...&0\\
	0&\mathcal{B^\dagger}&\mathcal{A}&\mathcal{B}&... &0\\
	0&0&\mathcal{B^\dagger}&\mathcal{A}&...&0 \\
	\vdots&\vdots&\vdots&\vdots&\ddots&\mathcal{B}\\
	\mathcal{B^\dagger} &0&0&0&\mathcal{B^\dagger}&\mathcal{A}
\end{pmatrix},\label{3. 1. 3}
\end{equation} 
\begin{align}
	\mathcal{A} &= \begin{pmatrix}
		0 & v_\downarrow & v_\uparrow & 0 \\
		v_\downarrow & 0 & 0 & v_\uparrow \\
		v_\uparrow & 0 & 0 & v_\downarrow \\
		0 & v_\uparrow & v_\downarrow & 0 
	\end{pmatrix}, \label{4} \\
	\mathcal{A'} &= \begin{pmatrix}
		V_1 & v_\downarrow & v_\uparrow & 0 \\
		v_\downarrow & 0 & 0 & v_\uparrow \\
		v_\uparrow & 0 & 0 & v_\downarrow \\
		0 & v_\uparrow & v_\downarrow & V_1 
	\end{pmatrix}, \label{5} \\
	\mathcal{B} &= \begin{pmatrix}
		0 & 0 & w_\uparrow e^{-iK} & 0 \\
		w_\downarrow & 0 & 0 & w_\uparrow e^{-iK} \\
		0 & 0 & 0 & 0 \\
		0 & 0 & w_\downarrow & 0 
	\end{pmatrix}. \label{6}
\end{align}

 In  Eqs. (\ref{4}),    the non-diagonal elements in $\mathcal{A}$ represent the intracell hopping of two electrons between their respective unit cells. In  Eqs. (\ref{5}), the non-zero diagonal elements of the matrix $\mathcal{A'}$ represent the interactions between the two electrons. The matrix $\mathcal{A'}$ only appears in the subspace of  $r_ {\uparrow \downarrow}=0 $.  In  Eqs. (\ref{6}), the non-diagonal elements  represent the intercell hopping of two electrons between their adjacent unit cells.  Note that the terms involving $w_\uparrow$ with a phase factor of $K$ due to the Fourier transformation. The matrices $\mathcal{B}$, positioned in the upper right and lower left corners, correspond to periodic boundary conditions. When considering open boundary conditions, these two elements are zero.

\subsection{States in the energy spectrum}
When the parameters in the Hamiltonian (such as $N$,   $V_1$,  $v_\uparrow$,    $v_\downarrow$,    $w_\uparrow$,    $w_\downarrow$) are fixed,    the Hamiltonian can be exactly diagonalized to obtain the energy spectrum.  

To investigate the effects of interactions and tunneling parameters on the two-particle states,    we  primarily focus on the real-space energy spectrum with open boundary conditions (OBC) .  For OBC,    particles can occupy the edge lattice sites of the chain,    leading to the emergence of edge states. 

\begin{figure}[htbp]
	\centering
	\includegraphics[width=0.5\textwidth]{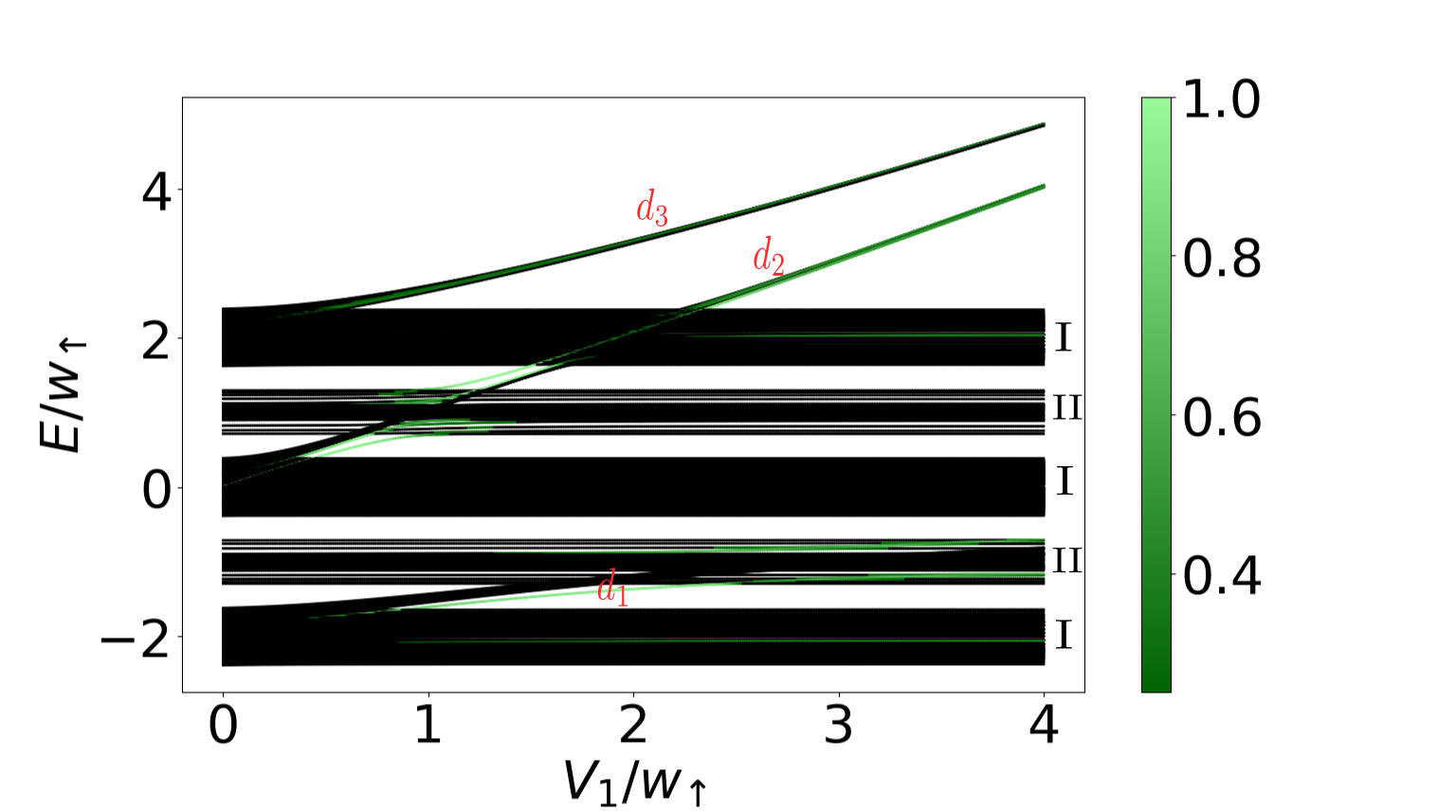}
	\captionsetup{justification=raggedright,    singlelinecheck=false}
	\caption{The energy spectrum as a function of the interaction \( V_1 \) for OBC with number of unit cells \( N = 12 \).  The intracell tunneling parameters are different: \( v_\uparrow = 0.1 w_\uparrow \) and \( v_\downarrow = 0.3 w_\downarrow \),    with \( w_\uparrow = w_\downarrow \). The color indicates the density of the two particles occupying the first and last unit cells simultaneously,    characterizing the edge localization of the states.  The broad black bands near energies \(\pm 2w_\uparrow\) and 0 correspond to the non-interacting bulk states of the two particles.  The sparse black bands near \(\pm w_\uparrow\) correspond to the non-interacting bulk-edge states.  The  narrow bands that vary with \( V_1 \) correspond to the correlated states of the two particles,    where the bright green narrow bands correspond to the correlated edge states and the black narrow bands correspond to the correlated bulk states. }
	\label{3-3}
\end{figure}

Initially, we consider the scenario where the intracell tunnelings $v_\uparrow$ and $v_\downarrow$ for two particles are distinct,    given by $v_\uparrow=0.1w_\uparrow$,   $v_\downarrow=0.3w_\uparrow$,    $w_\uparrow= w_\downarrow$.  The energy spectrum, which is dependent on the interaction $V_1$,    is presented in Fig. \ref{3-3}.  The two-particle states in the energy spectrum mainly include five types: (\romannumeral1) The non-interacting bulk state, characterized by particles behaving as free particles traversing in their respective SSH chains, occurs when both particles are in the bulk states without interactions. (\romannumeral2) The
 non-interacting bulk-edge states emerge under open boundary conditions (OBC), where particles can occupy edge lattice sites, and one particle resides in a bulk state while the other in an edge state without interaction between them.
 (\romannumeral3)The non-interacting edge-edge state characterizes a scenario where both free particles occupy edge states, a condition that is realized under open boundary conditions (OBC).  (\romannumeral4) The correlated bulk states describe the bound pairs formed by the two particles  due to the presence of interactions,    primarily located in the \dq{bulk} lattice sites.  (\romannumeral5) The correlated edge states describe the bound pairs of particles formed by the two particles through interaction $V_1$,    primarily located at the \dq{edge} lattice sites.     

Next, we will introduce the energy ranges of the five states separately. The energy of the non-interacting bulk state is composed of the superposition of the energy ranges of the single-particle bulk states. From the dispersion relation of the single-particle SSH model $E(k) = \pm \sqrt{v^2 + w^2 + 2vw\cos k}$ \cite{asboth_short_2016},    the energy range of the upper and lower bulk states for a single particle is $[-|v+w|,   -|v-w|]\cup [|v-w|,   |v+w|]$.  Therefore,    the bulk energy superposition of the two spin particles constitutes the  type \uppercase\expandafter{\romannumeral1} broad black energy bands near \(\pm 2w_\uparrow\) and 0 in the Fig. \ref{3-3}. 

 Given that two particles possessing distinct spins can respectively occupy edge states, this results in the emergence of two categories of non-interacting bulk-edge states. The particle in the edge state contributes no energy.  Thus,   the bandwidth of the non-interacting bulk-edge states is governed by the larger of the two single-particle bulk energy ranges.  This is represented by the type \uppercase\expandafter{\romannumeral2} sparse black energy bands near \(\pm w_\uparrow\) in the Fig. \ref{3-3}.

\begin{figure}[tpb]
	\centering
	\includegraphics[width=0.5\textwidth]{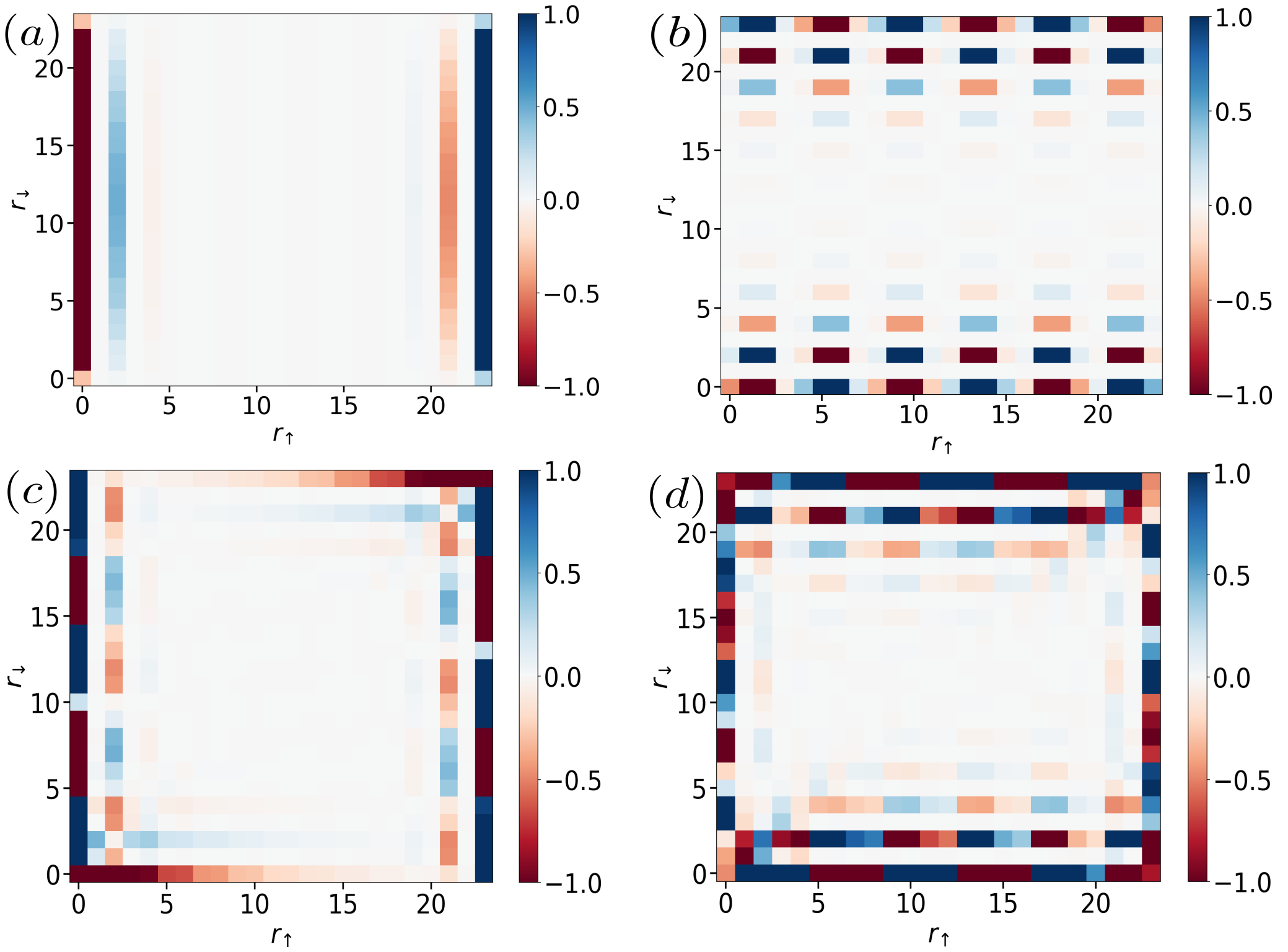}
	\captionsetup{justification=raggedright,    singlelinecheck=false}
	\caption{The wavefunction of the two types of non-interacting bulk-edge states, for $V_1=0$ and $V_1=1. 5w_\uparrow$,  with  other parameters are the same as in Fig.  \ref{3-3}.  Within the energy bands $[0. 7,   1. 3]w_\uparrow$ and $[-1. 3,   -0. 7]w_\uparrow$,    the states corresponds to the non-interacting two-particle bulk-edge states.  First row $V_1=0$: $(a)$ The edge state of an electron with spin up at energy $E\sim1. 29w_\uparrow$. $(b)$ The edge state of an electron with spin down at energy $E\sim1. 0w_\uparrow$. Second row $V_1=1. 5w_\uparrow$: $(c)$ The edge state of an electron with spin up at energy $E\sim1. 1w_\uparrow$.  $(d)$ The edge state of an electron with spin down at energy $E\sim0. 99w_\uparrow$. After adding the interaction $V_1 $,    the two types of edge states mix with each other.  For clarify,    the wavefunctions are amplified by $L$ times in the plot. }
	\label{3-4}
\end{figure}

The wavefunctions corresponding to the two types of non-interacting bulk-edge states are illustrated for two scenarios: $V_1=0$ and $V_1=1. 5w_\uparrow$,    as shown in Fig. \ref{3-4}. In contrast to the non-interacting scenario,    the presence of $V_1$ induces overlap between the edge-state wavefunctions of the two particles.  However,    the interaction does not correlate the dynamics of the two particles for these non-interacting bulk-edge states. Within the wave function diagram, the distribution of spin-up particles is plotted along the $x$-axis, and spin-down particles along the $y$-axis.  For the single-particle SSH model, occupation of solely the edge sites denotes an edge state, whereas occupation of all sites signifies a bulk state\cite{asboth_short_2016}. Under these conditions,    the decay length of the particle in the edge state still satisfies the single-particle decay length formula $\xi = \frac{1}{\log(w/v)}$\cite{asboth_short_2016}.  Since  
$\frac{w_\downarrow}{v_\downarrow} >\frac{w_\uparrow}{v_\uparrow} $,    the edge-state decay length of the spin-down particle is longer spanning approximately four unit cells, while the decay length of the spin-up particle is relatively short spanning approximately two unit cells (see Fig. \ref{3-4} (a)-(d)).

When both free particles occupy edge states, the energy of each edge state is zero. Consequently, the energy of non-interacting edge-edge state is also zero, representing a zero-energy state. In Fig. \ref{3-4}, the zero-energy state is situated within the band range of type \uppercase\expandafter{\romannumeral1}.

The energy spectrum also includes three clusters of narrow bands that depend on $V_1$.  These clusters correspond to correlated two-particle states,    influenced by interactions and exhibiting properties similar to the three intracell bound states described in \cite{di_liberto_two-body_2016}.  We label these states from lowest to highest energy as $d_1$, $d_2$ and $d_3$.  Detailed investigations were performed on $d_3$ and $d_2$,    as $d_1$  merges with other bands as $V_1$ increases. 

Correlated two-particle states can be categorized into two types: correlated bulk states and correlated edge states.  
Correlated bulk states describe the bound pairs formed by the two particles  primarily located in the \dq{bulk} lattice sites and occupying energies associated with bulk states. These states are depicted as black areas in the energy spectrum. And these states are considered topologically trivial. 

Correlated edge states describe the bound pairs of particles formed primarily located at the \dq{edge} lattice sites.  The probability of two particles occupying lattice sites decreases exponentially from the edge sites towards the central lattice sites. Within the energy spectrum, green denotes the edge state that arises when two particles are concurrently positioned at the edge. The utilization of single-particle topological invariants, such as the Zak phase\cite{di_liberto_two-body_2016}, is inadequate for describing two-body topological phases. Therefore, the presence of edge states is adopted as a diagnostic criterion for identifying the topological state. The green correlated edge states represent topologically non-trivial states.

\begin{figure}[tbp]
	\centering
	\includegraphics[width=0.5\textwidth]{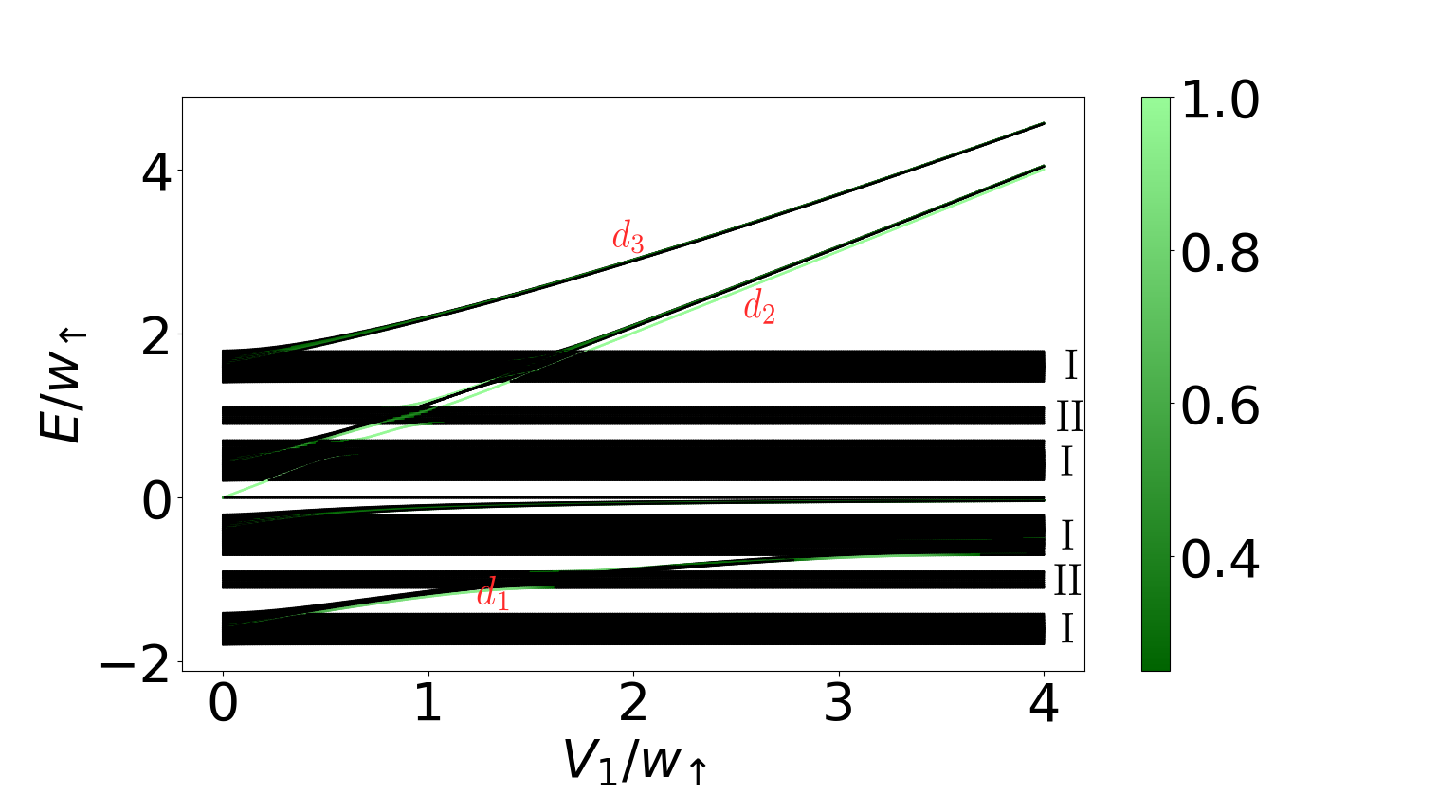}
	\captionsetup{justification=raggedright,    singlelinecheck=false}
	\caption{The energy spectrum as a function of the interaction \( V_1 \) for OBC with \( N = 12 \).  The intercell tunneling for two particles are different, $v_\uparrow=v_\downarrow=0. 1w_\downarrow$, $w_\uparrow = 0. 6w_\downarrow$. The green color indicates the density of the two particles occupying the first and last unit cells,    characterizing the edge localization of the states. }
	\label{3-5}
\end{figure}

Next,    consider the case where the intercell tunnelings $w_\uparrow$ and $w_\downarrow$ for two particles are different,  given by $v_\uparrow=v_\downarrow=0. 1w_\downarrow$,    $w_\uparrow = 0. 6w_\downarrow$, as shown in Fig. \ref{3-5}. The five types of states (the non-interacting bulk state, the
non-interacting bulk-edge states, the
non-interacting edge-edge states, the correlated bulk states and the correlated edge states) in the energy spectrum remain unchanged.  However,    the  range of non-interacting bulk-edge states and non-interacting bulk states have changed due to the variation in the energy range of single particle bulk states caused by the changes in $v $'s and $w $'s. In the energy gap between the two central type \uppercase\expandafter{\romannumeral1} energy bands in the middle of the four type \uppercase\expandafter{\romannumeral1} energy bands, there exists a non-interacting edge-edge state at zero energy and a bound state  whose energy reduces to zero with increasing $V_1 $. The non-interacting edge-edge state and the bound state are covered by the type \uppercase\expandafter{\romannumeral1} band in Fig.  \ref{3-3}, but they emerge clearly in Fig. \ref{3-5} . 

\begin{figure}[tbp]
	\centering
\includegraphics[width=0.5\textwidth]{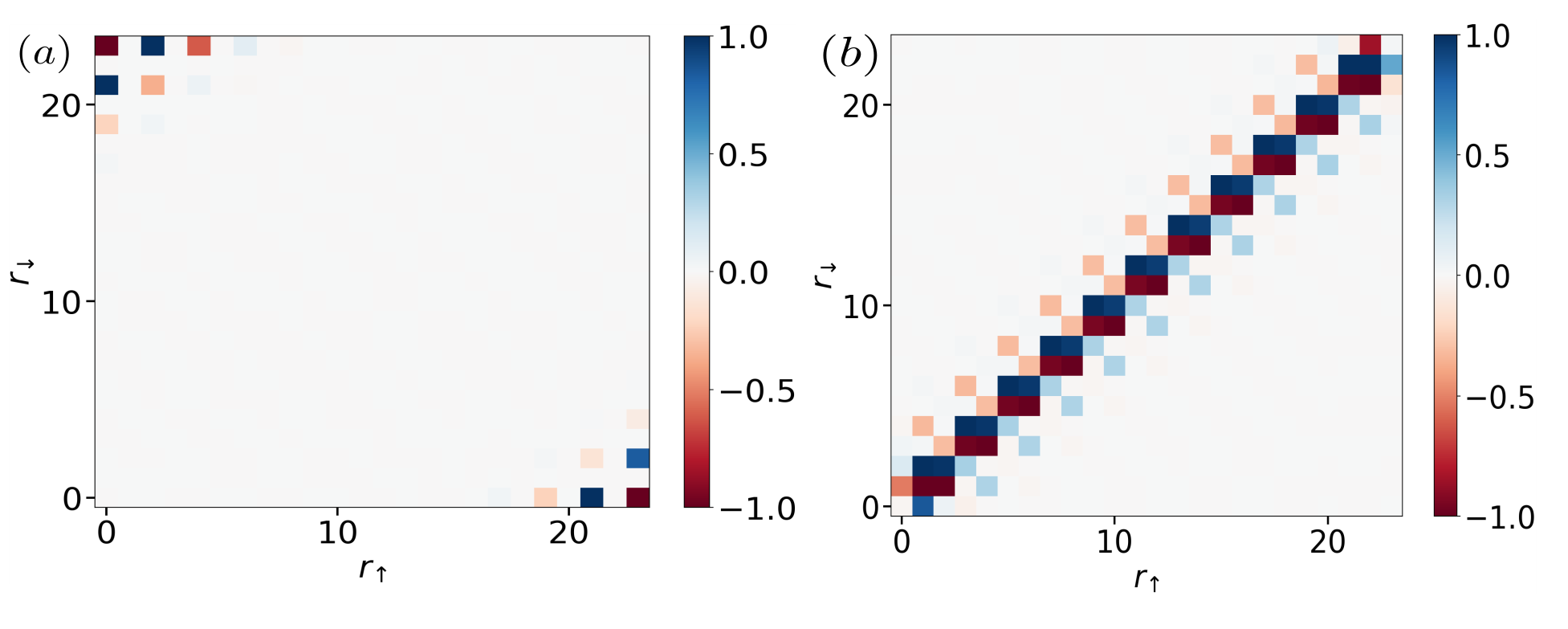}
	\captionsetup{justification=raggedright, singlelinecheck=false}
	\caption{The wave function distributions for two states with energy near zero are depicted,  when $V_1 $
		is set to $2w_\uparrow$.  (a) The degenerate non-interacting edge-edge state wave functions at $E=0$, where two particles are occupying the terminal lattice sites on opposing edges. (b) The extra bound state wave function with $E\sim-0.08w_\uparrow$, where the two particles primarily occupy the same lattice site and adjacent sites.  }
	\label{3-17}
\end{figure}

When $V_1 $
is set to $2w_\uparrow$, the typical wave functions of the non-interacting edge-edge state and the bound state are depicted in Fig. \ref {3-17}. The non-interacting edge state exhibits double degeneracy, with the two particles residing at opposing edge lattice sites. Consequently, the wave function is predominantly localized in the upper left and lower right corner. The extra bound state appears due to presence of interaction $V_1 $. The wave function indicates that the two particles are predominantly located at the same lattice site and adjacent lattice site.

\begin{figure}[htbp]
	\centering
    \includegraphics[width=0.5\textwidth]{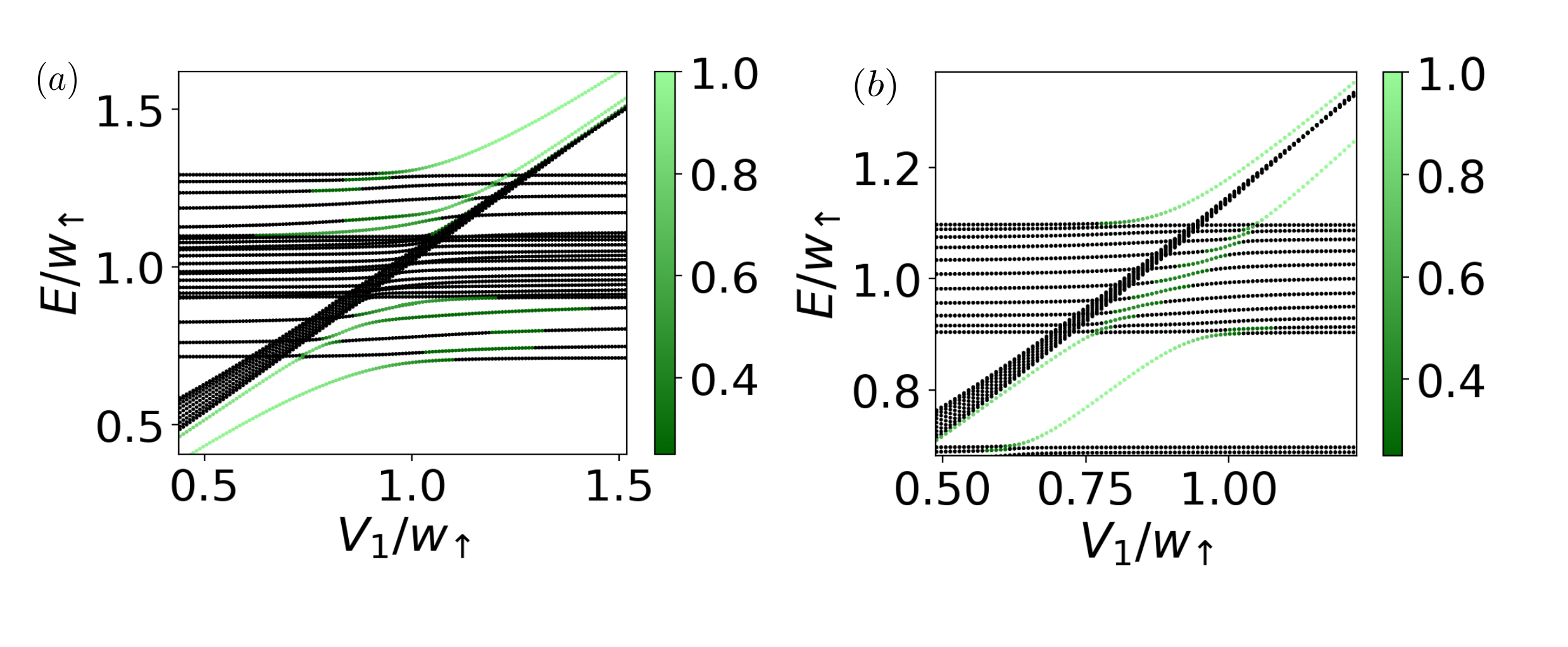}
	\captionsetup{justification=raggedright,    singlelinecheck=false}
	\caption{The energy spectrum of the $d_2$ band near $V_1=w_\uparrow$ under OBC,    where $N=12$. (a) For different intracell tunnelings $v_\uparrow$ and $v_\downarrow$ for particles are distinct, given by  $v_\uparrow=0.1w_\uparrow$,   $v_\downarrow=0.3 w_\uparrow$,   $w_\uparrow=w_\downarrow$,    the strongly localized edge state splits into two distinct strongly localized states. (b) For different intercell tunnelings,    $v_\uparrow=v_\downarrow=0. 1w_\downarrow$,   $w_\uparrow=0. 6w_\downarrow$,    the strongly localized edge state splits into two distinct strongly localized states,    which are further shifted to the right compared to the results in (b). }
	\label{3-6}
\end{figure}

We subsequently investigated the changes of the $d_2$ correlated edge states in Fig.  \ref{3-3} and Fig. \ref{3-5}  as \( V_1 \) increases.  Initially,    by focusing on the energy spectrum region near $V_1=w_\uparrow$
for the $d_2$ state with two different sets of spin tunneling parameters as shown in Fig. \ref{3-6}.   we observed that,    unlike the case of identical spin tunneling ($v_\uparrow=v_\downarrow$,   $w_\uparrow=w_\downarrow$ as shown in Appendix B),    when spin tunneling is distinct for either intracell tunneling ($v_\uparrow=0. 1w_\uparrow$,   $v_\downarrow=0. 3w_\uparrow$ as shown in Fig.  \ref{3-6}(a) ) or intercell tunneling ($v_\uparrow=v_\downarrow$,   $w_\uparrow=0. 6w_\downarrow$ as shown in Fig.  \ref{3-6}(b)) is varied,    the edge state energy bands lift the degeneracies.  Additionally,    we found that when spin tunnelings for two spins are distinct, edge states (highlighted in bright green) and bulk states (indicated in black) are energetically separated,    whereas when spin tunnelings for different spins are identical, edge states and bulk states overlap in energy. 

\begin{figure}[htbp]
	\centering
    \includegraphics[width=0.9\linewidth]{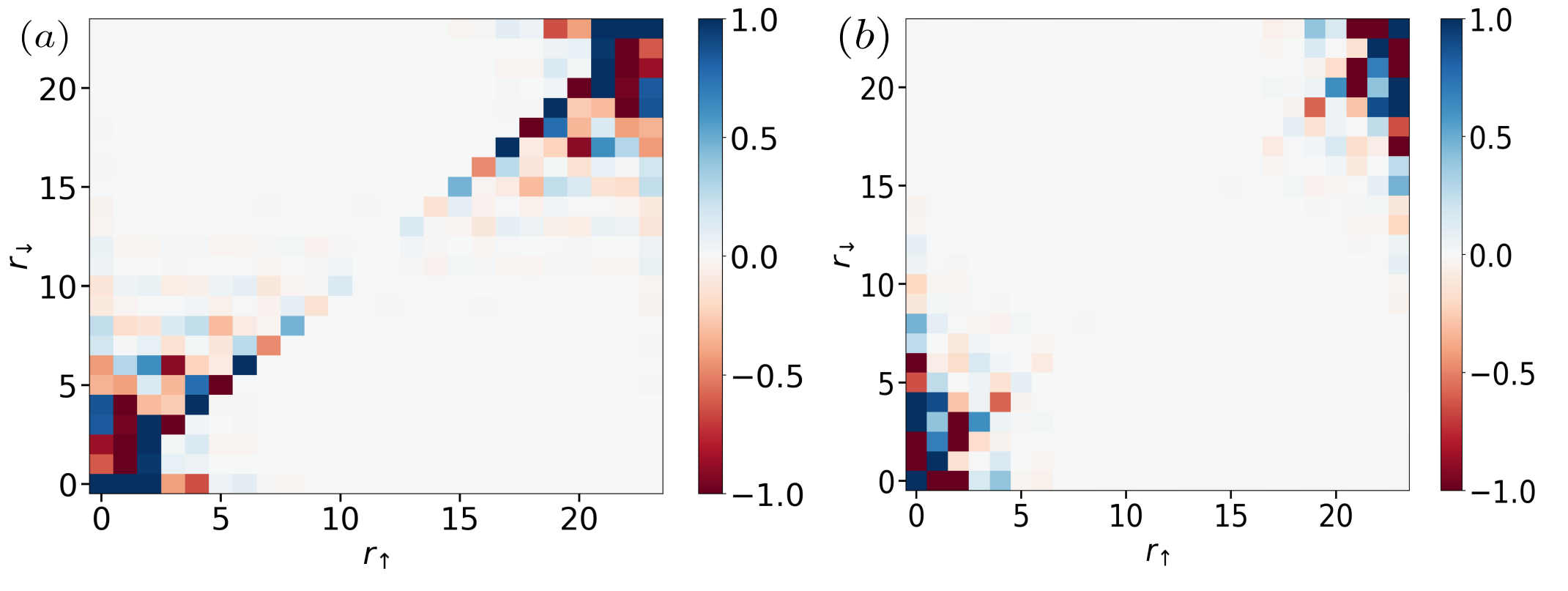}
\captionsetup{justification=raggedright,    singlelinecheck=false}
	\caption{When $V_1 = 0.6 w_\uparrow$, the intracell tunneling parameters vary, whereas the remaining parameters align with those in Figure \ref{3-6}. This setup illustrates the distribution of strongly localized correlated edge state wave functions. (a) For identical intracell tunneling of two electrons $v_\uparrow = v_\downarrow = 0.1 w_\uparrow$ and $w_\uparrow=w_\downarrow$, the correlated edge state at  energy  $E\sim0.57w_\uparrow$. (b) and (c) feature distinct intracell tunneling of the two electrons, with $v_\uparrow = 0.1 w_\uparrow$, $v_\downarrow = 0.3 w_\downarrow$ and $w_\uparrow=w_\downarrow$. The correlated edge states at energies of $E\sim0.6w_\uparrow$
		and $E\sim0.5w_\uparrow$, respectively.}
	\label{fig:3-9}
\end{figure}

In the vicinity of $V_1=w_\uparrow$
within the energy spectrum, we selected $V_1=0.6w_\uparrow$
and depicted the wave function distributions corresponding to two strongly localized edge states, which are presented in Fig.  \ref{3-6}(a) , and illustrated in Fig.   \ref{fig:3-9}. In contrast to the scenario where the wave function images for identical intercell tunneling are symmetric about the diagonal (see Appendix B), the wave function images for different intercell tunneling do not exhibit strict symmetry along the diagonal, and the decay length of spin-down particles is extended (Fig.  \ref{fig:3-9}(a)(b)).


\begin{figure}[htbp]
	\centering
    \includegraphics[width=0.9\linewidth]{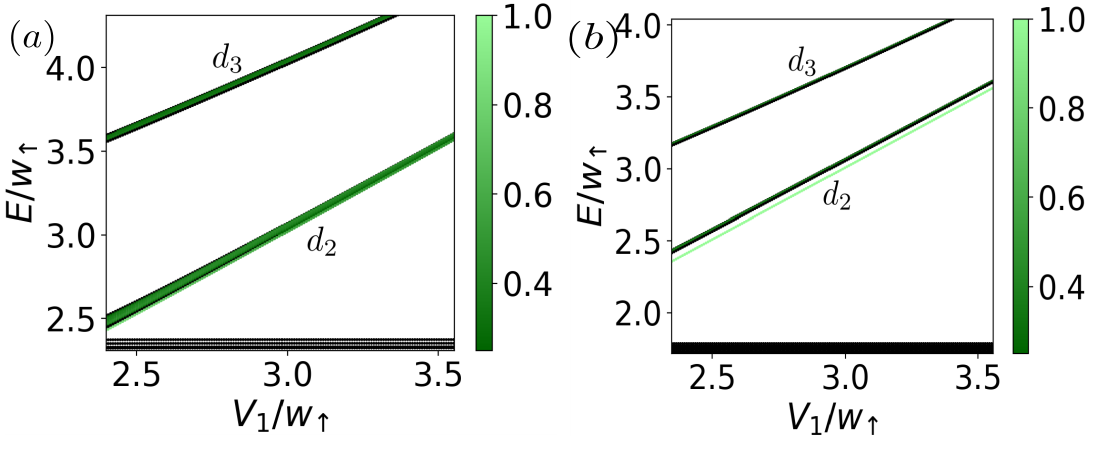}
	\captionsetup{justification=raggedright,    singlelinecheck=false}
	\caption{The energy spectrum of $d_3$ and $d_2$ near $V_1=3w_\uparrow$ under OBC with $N=12$.  The higher energy tilted band is $d_3$,    and the lower energy tilted band is $d_2$. (a) For different intracell tunnelings,   $v_\uparrow=0. 1w_\uparrow$,   $v_\downarrow=0. 3w_\uparrow$,   $w_\uparrow=w_\downarrow$,    $d_2$ contains mostly correlated edge states and a small fraction of correlated bulk states,    whereas $d_3$ is dominated by correlated bulk states.  Thus,  $d_2$ is green lighter,    and $d_3$ is darker. (b) For different intercell tunnelings,  $v_\uparrow=v_\downarrow$, $w_\uparrow=0. 6w_\downarrow$,      $d_3$ is almost entirely consist of correlated bulk states,  whereas $d_2$ exhibits a energy gap between correlated bulk states and correlated edge states. }
	\label{3-7}
\end{figure}

Subsequently, we concentrate on the energy spectrum region near $V_1=3w_\uparrow$
for the $d_2$ state, considering distinct intracell ($v_\uparrow$,$v_\downarrow$) and intercell tunneling ($w_\uparrow$,$w_\downarrow$) parameters, as depicted in Fig. \ref{3-7}(a) and (b), respectively. It is observed that energy gap between the correlated bulk state $d_2$ and the correlated edge $d_2$  decreases  with increasing $V_1$.  As indicated by the energy spectrum colors in Fig.  \ref{3-7}(a),    $d_3$ is primarily composed of correlated bulk states (characterized by dark green in the spectrum),    whereas $d_2$ is a combination of correlated bulk and edge states (with both black and green colors in the spectrum,    and green being dominant at higher $V_1$). As shown in Fig.  \ref{3-7}(b),    $d_3$ is almost entirely consist of correlated bulk states,    whereas in $d_2$,    the strongly localized correlated edge states exhibit a noticeable energy shift from the correlated bulk states,    resulting in an energy gap. 

\begin{figure}[htbp]
	\centering
    \includegraphics[width=0.9\linewidth]{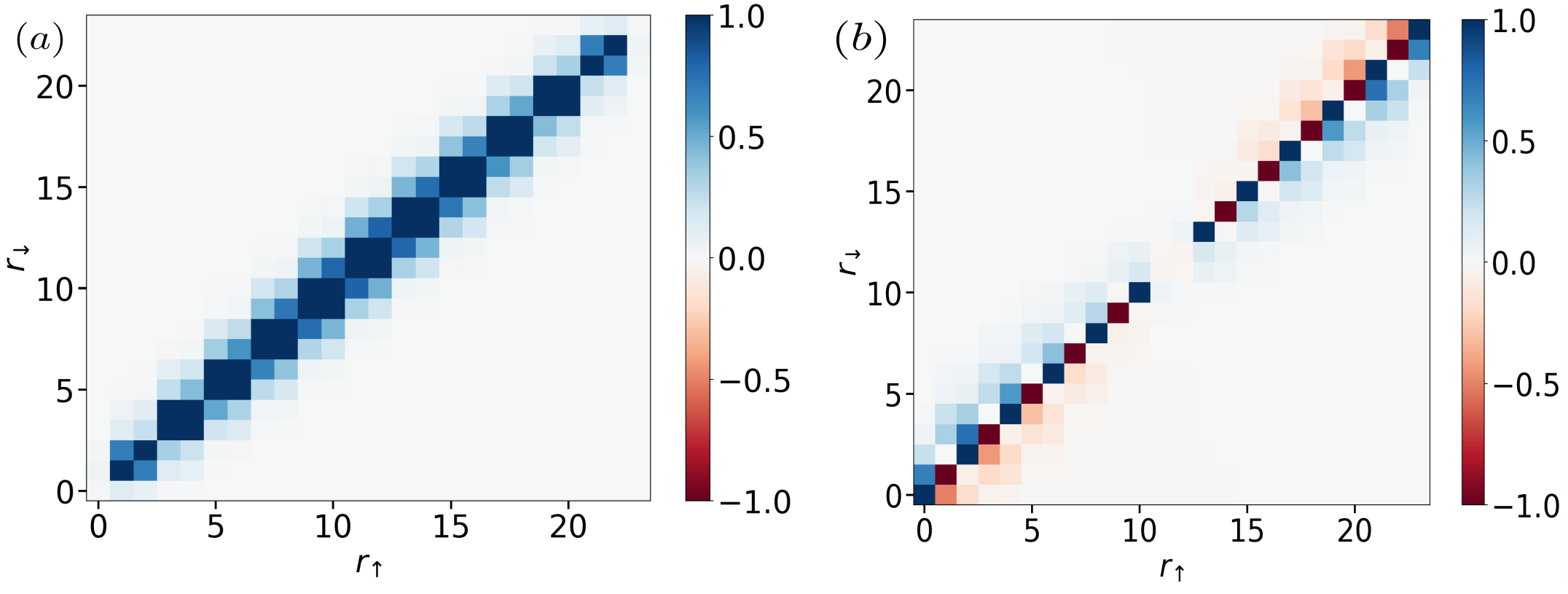}
	\captionsetup{justification=raggedright,    singlelinecheck=false}
	\caption{The wavefunction distributions of the two-particle correlated states $d_3$ and $d_2$ at interaction $V_1=3w_\uparrow$ are shown in Fig.  \ref{3-7}(a).  (a) The two-particle correlated bulk state in $d_1$, at $E\sim4. 05w_\uparrow$. (b)  The two-particle correlated edge state in $d_2$ at $E\sim3. 01w_\uparrow$ shows an  decay from the chain ends towards the center. }
	\label{3-8}
\end{figure}

To validate our conclusions,    we generated the wavefunction distributions for the $d_3$
and $d_2$
states as depicted in Fig. \ref{3-7}(a),    which are presented in Fig. \ref{3-8}.  Specifically,    (a) illustrate the correlated bulk states of $d_3$,    whereas (b) highlights the correlated edge state of $d_2$. The occupancy probabilities  at the central sites decay to zero for the correlated edge state of $d_2$.

Next,    we performed a detailed analysis on the impact of interactions and differences in spin-dependent tunneling intensities ($|v_\uparrow-v_\downarrow|$ or $|w_\uparrow- w_\downarrow|$) on the decay length and dynamical behavior of the correlated edge states $d_2$.

\section{Decay length of correlated edge state}

In order to thoroughly comprehend the factors that affect the decay length,    we have qualitatively investigated the impact of interactions and the spin tunneling differences $|v_\uparrow-v_\downarrow|$ or $|w_\uparrow- w_\downarrow|$ on the decay length of the correlated edge state.

\begin{figure}[htbp]
	\centering
    \includegraphics[width=0.9\linewidth]{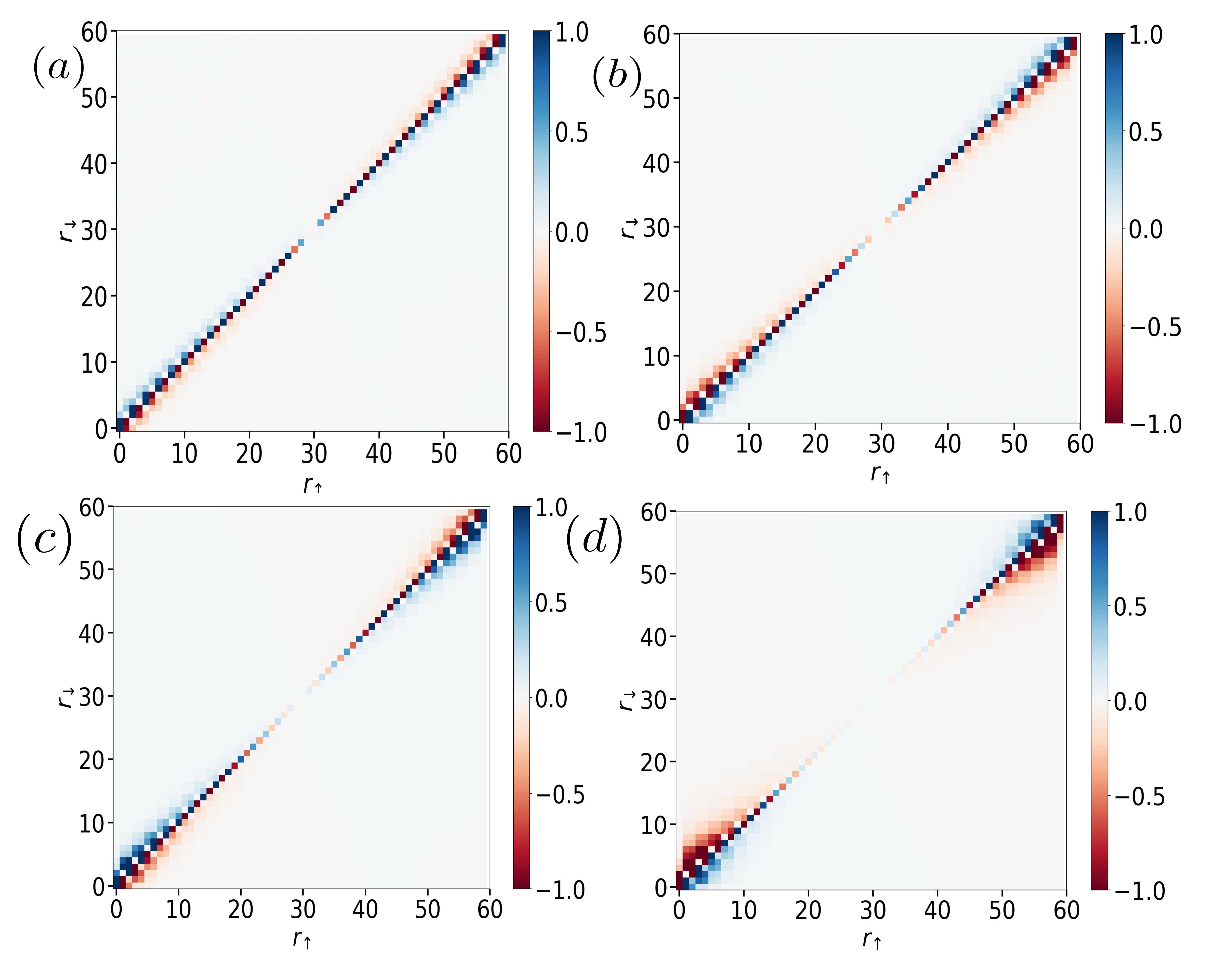}
	\captionsetup{justification=raggedright,    singlelinecheck=false}
	\caption{The wavefunctions of the correlated two-particle edge states at different $V_1$ values.  $N=30$,  $v_\uparrow=0. 1w_\uparrow$,    $v_\downarrow=0. 3w_\downarrow$,    $w_\uparrow=w_\downarrow$. 
		(a) to (d) depict the wave functions of the same edge state for different $V_1$ values,    corresponding to interaction strengths of $V_1=3. 5w_\uparrow$,    $V_1=3w_\uparrow$,    $V_1=2. 75w_\uparrow$,    $V_1=2. 5w_\uparrow$. 
		When $V_1=2. 5w_\uparrow$,    the decay length exhibits an exponential decay,    whereas when $V_1=3. 5w_\uparrow$,    the decay length increases and does not follow an exponential decay. }
	\label{fig:3-11}
\end{figure}

\begin{figure}
	\centering
	\includegraphics[width=8.00cm,   height=6cm]{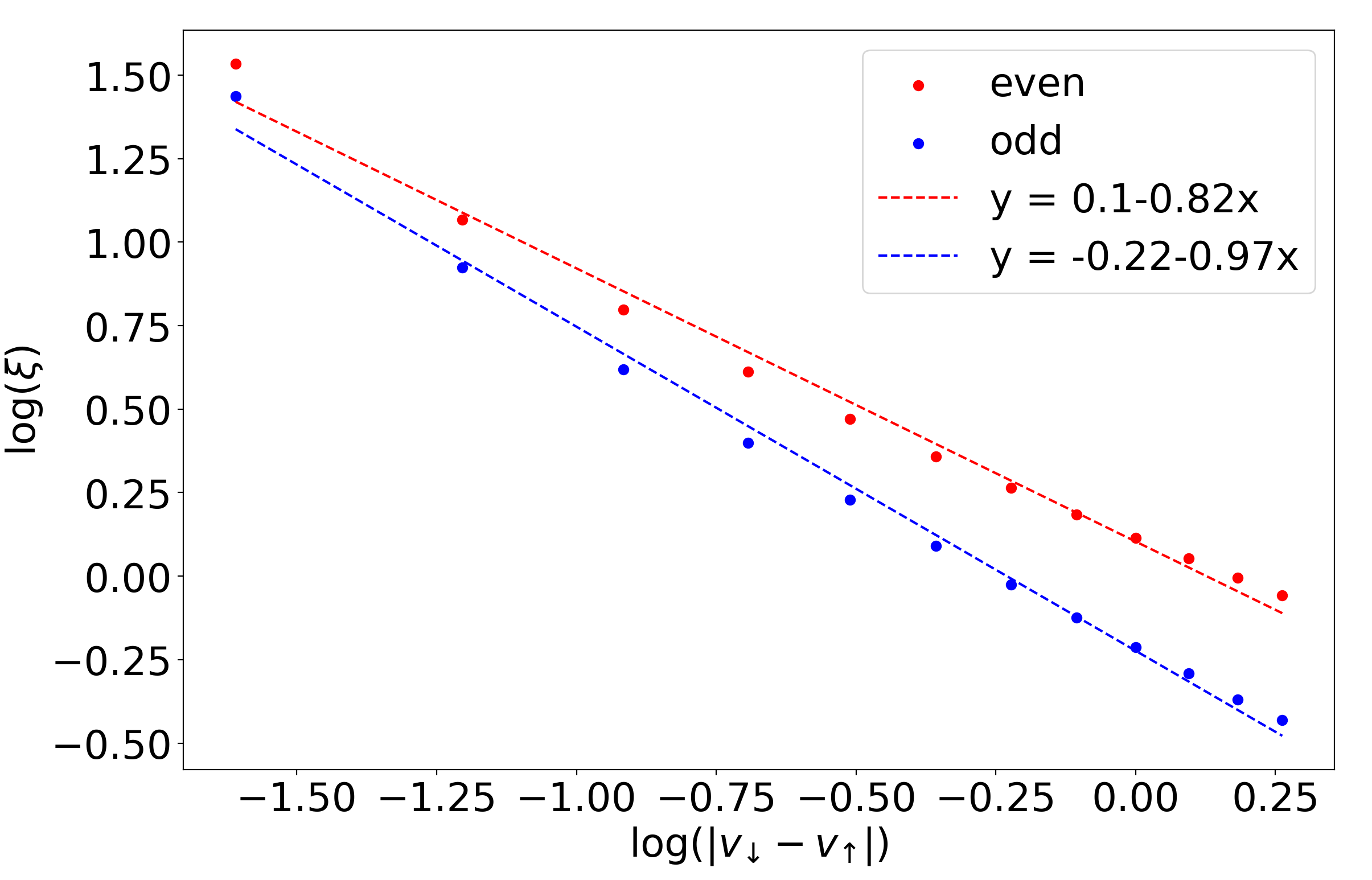}
	\captionsetup{justification=raggedright,    singlelinecheck=false}
	\caption{%
		The fitting of the dependence of the decay length of wavefunction  on  $| v_\uparrow-v_\downarrow | $ at even and odd lattice sites associated with the $d_2$ edge states.   $N=30$,   $V_1=3w_\uparrow$,     $v_\uparrow=0. 1w_\downarrow$,    $v_\downarrow$ increases uniformly from $0. 3w_\downarrow $ to $1. 4w_ \downarrow $.  The fitting value of decay length follows a power-law  decay with the increase of $| v_\uparrow-v_\downarrow | $.  The larger the $| v_\uparrow-v_\downarrow| $,    the shorter the decay length of the correlated edge state. }
	\label{3-12}
\end{figure}

\begin{figure}
	\centering
    \includegraphics[width=8.00cm,   height=10cm]{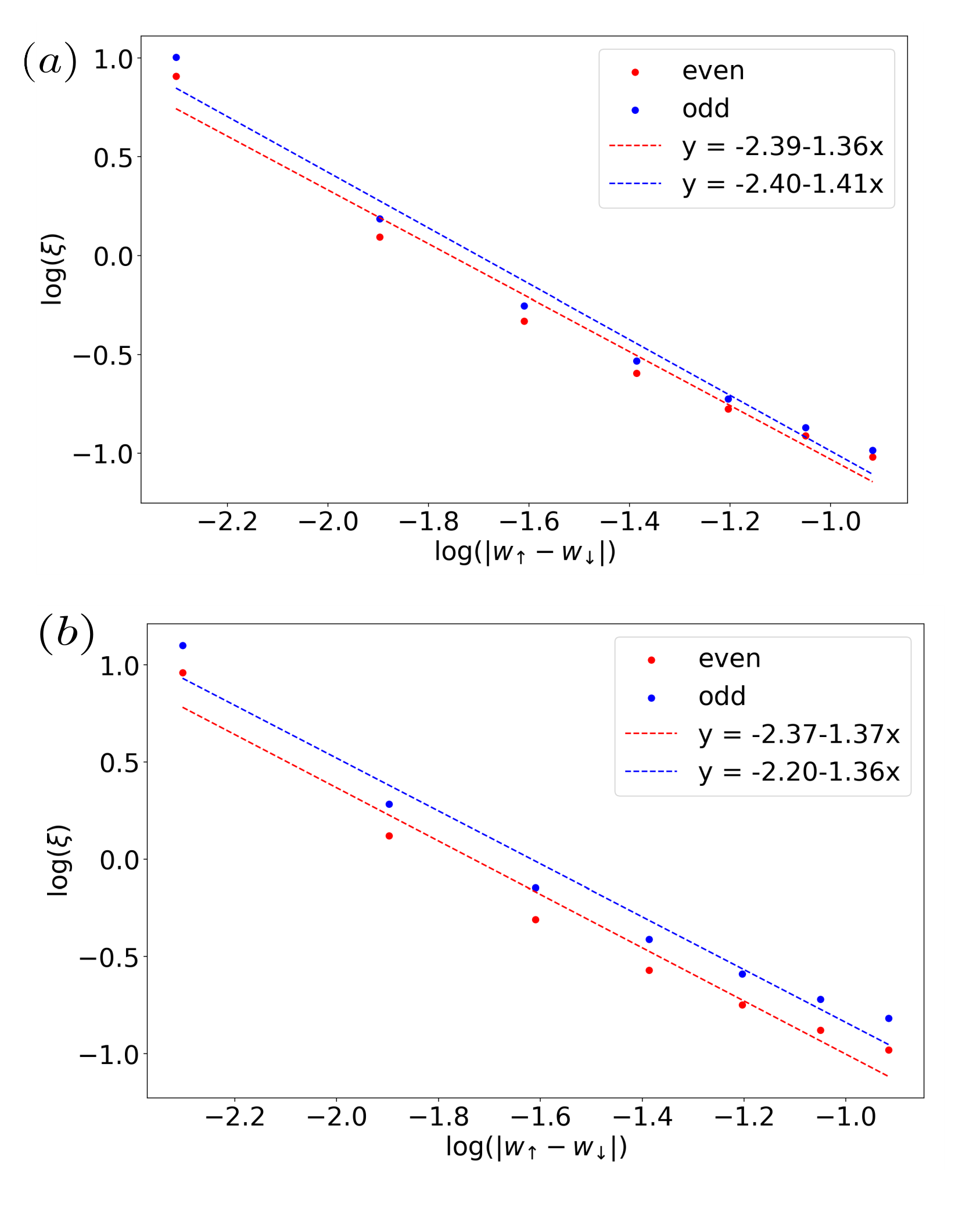}
	\captionsetup{justification=raggedright,    singlelinecheck=false}
	\caption{The dots are from the numerical simulatin of the decay length for the wavefunction at the even and the odd sites, and the straight lines are from the fittings of the dependence on  $| w_\uparrow-w_\downarrow | $ of the decay length of wavefunctions at even and odd lattice sites associated with the edge states of two particles.  
		$N=15$,   $V_1=3w_\downarrow$,  $v_\uparrow=v_\downarrow=0. 1w_\downarrow$.  We uniformly reduce $w_\uparrow$
		from $0. 9w_\downarrow$
		to $0. 6w_\downarrow$
		in (a) and uniformly increase $w_\uparrow$
		from $1. 1w_\downarrow$
		to $1. 4w_\downarrow$
		in (b).  The fitting value of decay length follows a power-law  decay with the increase of $| w_\uparrow-w_\downarrow | $.  A larger $|w_\uparrow-w_\downarrow|$ corresponds to a shorter decay length of the edge state. 
	}
	\label{3-13}
\end{figure}
We first investigated the effect of the interaction $V_1$
on the decay length of the correlated edge state $d_2$. 
We chose the same correlated edge state and presented the probability amplitudes of the wave function along the diagonal for four different interaction strengths,    as depicted in Fig.  \ref{fig:3-11}. (a) to (d) depict the wave functions of the same edge state for different $V_1$ values,    corresponding to interaction strengths of $V_1=3. 5w_\uparrow$,    $V_1=3w_\uparrow$,    $V_1=2. 75w_\uparrow$,    $V_1=2. 5w_\uparrow$. 	We show the projections along the diagonal $x=y$ of (a) to (d) in Appendix C. The diagonal indicates that the two particles are aligned at the same lattice site.  The probability amplitudes decrease  from the ends of the chain to the central lattice site,    and the decay length increases as $V_1$ increases. Note that, when $V_1$ is relatively small,    the decay length follows an exponential decay formula $y=ae^{-(x-1)/b}+c$, where $x$ represents the position of sites, $b$ represents the decay length of state, $a$ and $c$ are fitting parameters. On the other hand, when $V_1$ is large, the decay length increases, and due to the finite size effect, the decay trend no longer follows exponential decay.

We subsequently investigated how varying spin tunneling  impact the decay length of correlated edge states $d_2$.  Initially,    we explored the effects of different spin intracell tunneling by fixing $N=30$,   $V_1=3w_\uparrow$,  $v_\uparrow=0. 1w_\uparrow$,   $w_\uparrow=w_\downarrow$.  We uniformly increased $v_\downarrow$ from $0. 3w_\uparrow$ to $1. 4w_\uparrow$ to alter the intracell tunneling difference $|v_\uparrow-v_\downarrow|$. By extracting the probability amplitude distribution of the edge state wave function along the diagonal and conducting numerical fitting,   
we discovered that the decay lengths $\xi$ of odd (even) sites and $|v_\uparrow-v_\downarrow|$ follow a power-law function decay $y=ax^{-b}$, where $x$ represents the position of sites,  $a$ and $b$ are fitting parameters. To more clearly demonstrate this relation,    we perform log-log plot of the decay length as a function of $|v_\uparrow-v_\downarrow|$. The result satisfies linear function  $ln(y)=ln(a)-bln(x)$  as illustrated in Fig.  \ref{3-12}. 

 The red and blue dots represent the fitting values of the decay lengths of even and odd sites along the diagonal $x=y $ of the correlated edge state wave function with a fixed parameter.  The fitting value of decay length follows a power-law  decay with the increase of $| v_\uparrow-v_\downarrow | $.  After taking logarithms for two types of decay lengths and $| v_\uparrow-v_\downarrow | $,    both fitting relations (red and blue dashed lines) exhibit a linear decay trend.  The larger the $| v_\uparrow-v_\downarrow| $,    the shorter the decay length of the correlated edge state.

Next,    we explored the effects of different spin intercell tunneling by fixing $N=15$,   $V_1=3w_\downarrow$, $v_\uparrow=v_\downarrow=0. 1w_\downarrow$,    we uniformly reduce $w_\uparrow$
from $0. 9w_\downarrow$
to $0. 6w_\downarrow$
or uniformly increase from $1. 1w_\downarrow$
to $1. 4w_\downarrow$ to alter the intercell tunneling difference $|w_\uparrow-w_\downarrow|$.  We found that the dependence of the decay lengths $\xi$ and $|w_\uparrow-w_\downarrow|$ also satisfy the power-law  $y=ax^{-b}$. And we equivalently convert this relation into linear functions $ln(y)=ln(a)-bln(x)$   as illustrated in Fig.  \ref{3-13}. The larger the $|w_\uparrow-w_\downarrow|$,    the shorter the decay length of the correlated edge state.

\section{Dynamics of correlated edge state}

The decay length of edge states is intrinsically linked to their dynamical behavior. We have investigated how interactions and spin-dependent intracell (intercell) tunneling affect the decay length of  correlated edge states $d_2$.  For a given edge state,    its dynamical behavior can be entirely different under different $V_1$ values and spin tunneling,    primarily due to changes in decay length. 
\begin{figure}[htbp]
	\centering
    \includegraphics[width=9.00cm,   height=12cm]{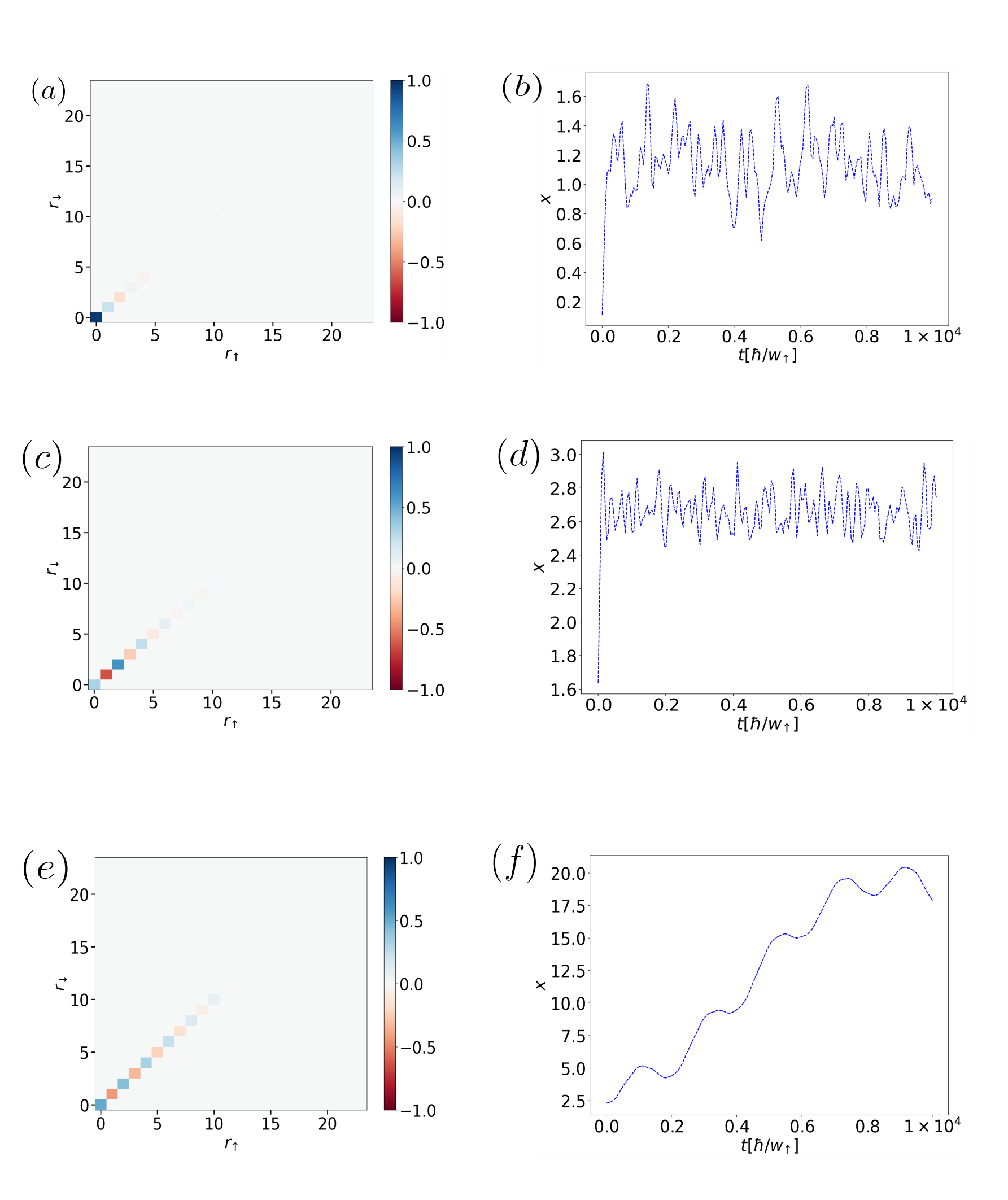}
	\newline 

	\captionsetup{justification=raggedright,    singlelinecheck=false}
	\caption{The wave functions of the correlated edge states $d_2$ for various $V_1$	values and the time-evolving expectation values of the coordinate $x$ after initialization.  First row: With $V_1=0. 6w_\uparrow$,   $E\sim0. 5w_\uparrow$,    the initialized wave function is localized within the first three unit cells at the edge.  Over time,    the two electrons are primarily localized near the middle of the first and second lattice sites,    maintaining localization.  Second row: With $V_1=0. 6w_\uparrow$,   $E\sim0. 6w_\uparrow$,    the initialized wave function is localized within the first four unit cells at the edge.  Over time,    the two electrons are primarily localized between the second and third lattice sites,    maintaining localization.  Third row: With $V_1=3w_\uparrow$,   $E\sim3w_\uparrow$,    the wave function initial decay length is relatively long. Over time, the electrons move along the chain and are not localized at the edge. }
	\label{3-14}
\end{figure}

First,    we explored   correlated edge states of $d_2$ with different intracell tunneling parameters $v_\uparrow=0. 1w_\uparrow$,   $v_\downarrow=0. 3w_\uparrow$   (see also Fig. \ref{3-6}(b)).  We projected their wave functions onto the diagonal $x=y$ and truncated them to half the chain length.  These truncated wave functions were normalized and used as the initial states for time evolution as depicted in Fig.  \ref{3-14}(a) and (c).  To characterize the dynamical behavior of the two correlated electrons,    we monitored the expectation value of their  coordinates $\langle x \rangle$.  The results indicated that by $t=1\times10^
4$,    particles in these two edge states were predominantly localized within three edge lattice sites,    demonstrating strong localization,    as depicted in Fig.  \ref{3-14}(b) and (d). 

However,    the same correlated edge state of $d_2$ at $V_1=3w_\uparrow$
exhibits a longer decay length and distinct dynamical behavior compared to the two correlated edge states at $V_1=0. 6w_\uparrow$.  Using the same method to prepare the initial state,    we observed its dynamical behavior,    as depicted in Fig. \ref{3-14} (e) and (f).  On the same time scale,    the two particles traverse the entire chain rather than remaining localized at the edges.  The particles undergo oscillatory motion along the chain. 

\begin{figure}
	\centering
    \includegraphics[width=9.00cm,   height=12cm]{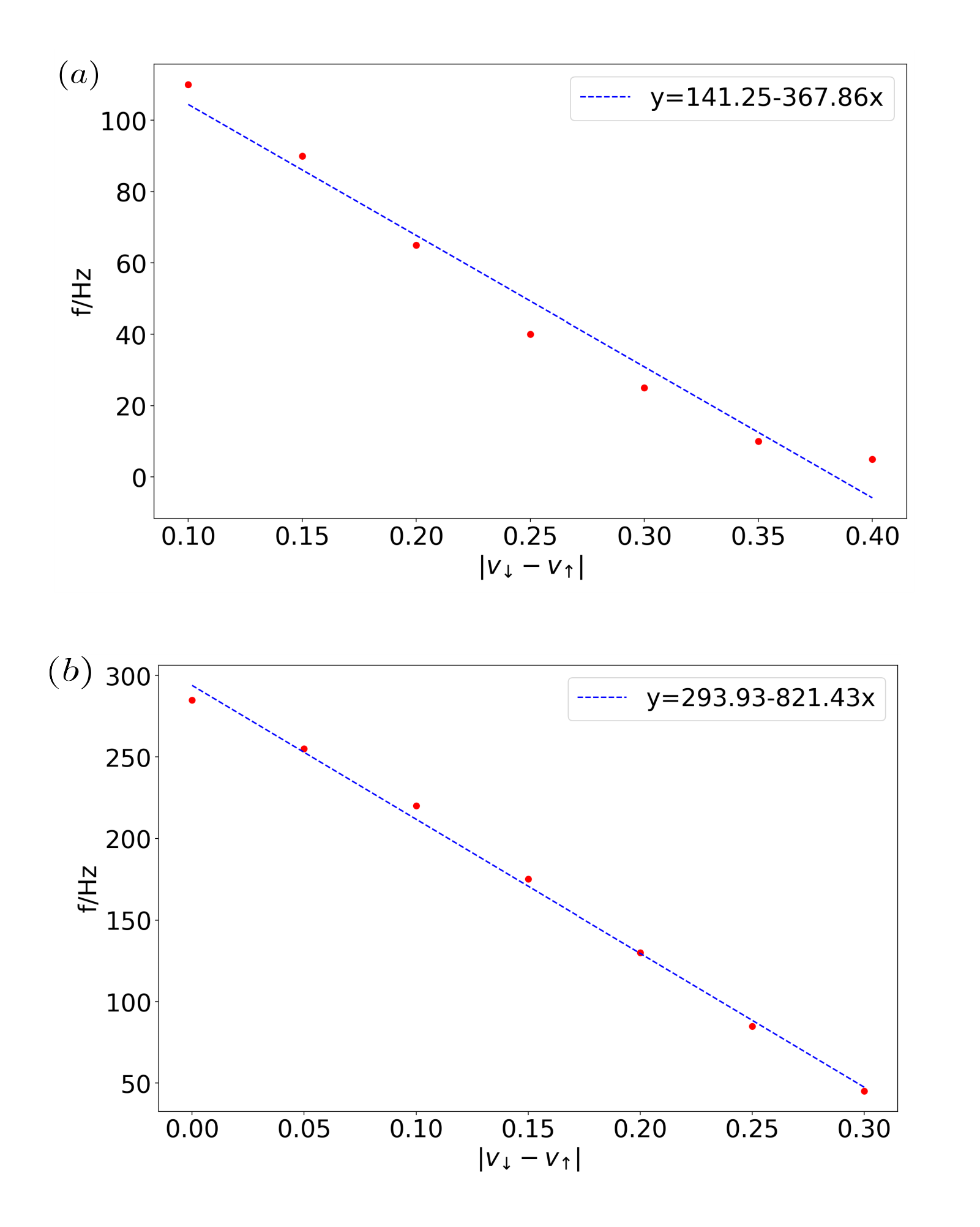}
	\captionsetup{justification=raggedright,    singlelinecheck=false}
	\caption{The relation between \(|v_\downarrow - v_\uparrow|\) and the oscillation frequency of the correlated edge states along the chain is investigated.  \(N=30\),    \(V_1=3w_\uparrow\),     \(w_\uparrow = w_\downarrow\).  (a) \(v_\uparrow = 0. 1w_\uparrow\) with \(v_\downarrow\) uniformly increased from \(0. 2w_\downarrow\) to \(0. 5w_\downarrow\).  (b) \(v_\downarrow = 0. 5w_\uparrow\) with \(v_\uparrow\) uniformly increased from \(0. 2w_\uparrow\) to \(0. 5w_\uparrow\).  The red dot indicates the fitted oscillation frequency of the two particles in the edge state,    and the fitting  (blue dashed line) shows that the oscillation frequency $f$ is proportional to $|v_\downarrow - v_\uparrow|$.  A smaller \(|v_\downarrow - v_\uparrow|\) corresponds to a higher oscillation frequency due to the longer decay length. }
	\label{3-16}
\end{figure}

By comparing the  correlated edge states of $d_2$ at different $V_1$ values,    we found that larger interactions not only increase the decay length of edge states but also induce significant particle oscillatory motion.  This observation is in accordance with the general rule that as the decay length increases,    the distance over which the particle propagates also increases. 

As  discussed previously,    when $V_1=3w_\uparrow$,    the decay length of the edge state exhibits a power-law  decay with the difference $|v_\downarrow-v_\uparrow|$. We assume that smaller differences result in longer decay lengths,    and particles moving along the chain would exhibit higher oscillation frequencies.  The linear fitting function depicted in Fig. \ref{3-16} supports our hypothesis. 

\begin{figure}
	\centering 
    \includegraphics[width=9.00cm,   height=8cm]{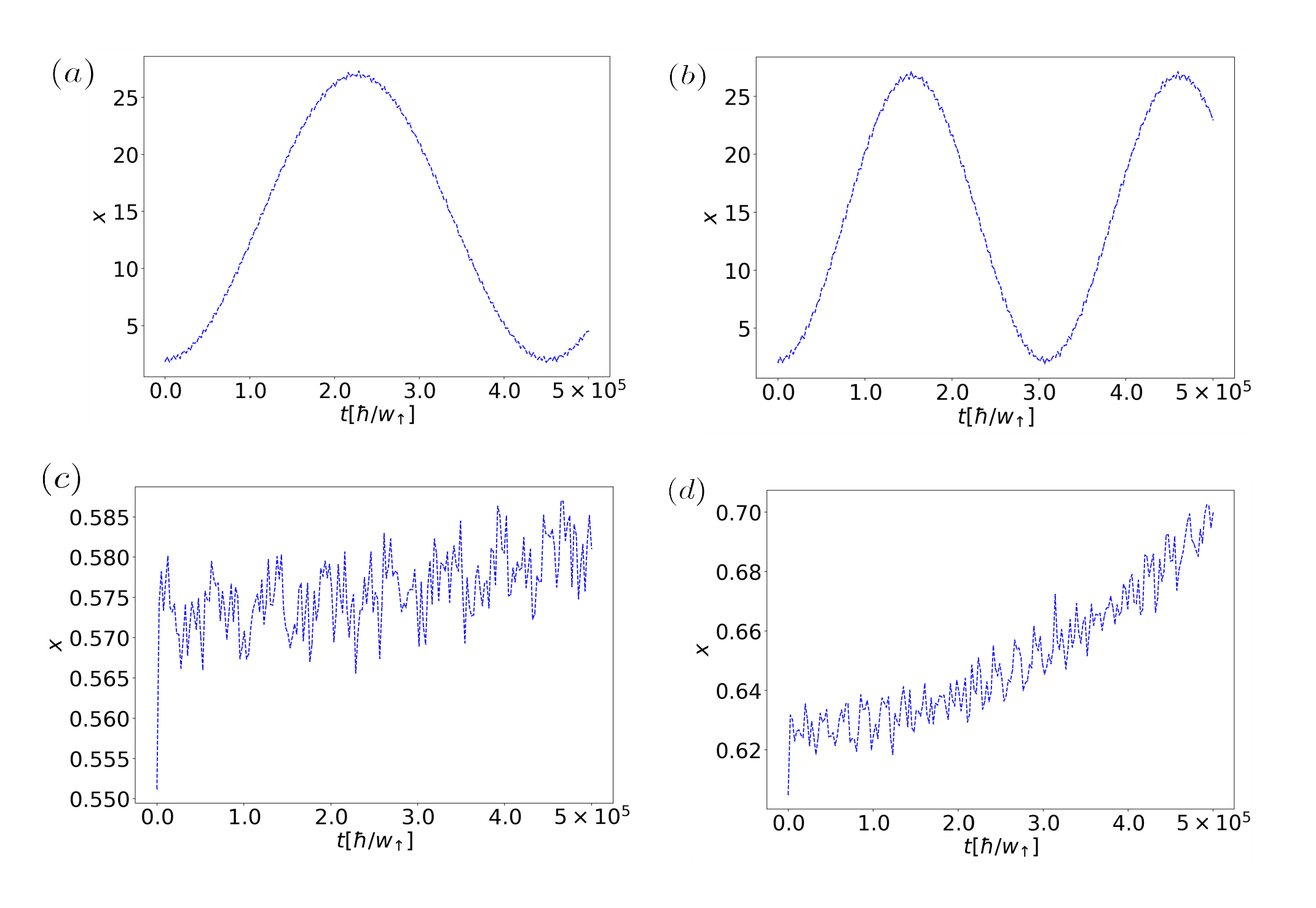}
	\captionsetup{justification=raggedright,    singlelinecheck=false}
	\caption{The relation between $|w_\uparrow-w_\downarrow|$ and the oscillatory motion of the correlated edge states associated with two particles along the chain is examined. \(N=15\),    \(V_1=3w_\downarrow\),     \(v_\uparrow = v_\downarrow=0.1w_\downarrow\). The first row features parameters fixed at $|w_\uparrow-w_\downarrow|=0.1$. (a)$w_\uparrow=0.9w_\downarrow$,    and (b) $w_\uparrow=1.1w_\downarrow$.  At this stage,    the two particles oscillate back and forth along the entire chain.  The second row features parameters fixed at $|w_\uparrow-w_\downarrow|=0.05$ (c) $w_\uparrow=0.85w_\downarrow$,    and (d)$w_\uparrow=1.15w_\downarrow$. In (c) and (d), the two particles are localized and oscillate within a single site at the end,    without traversing the chain. }
	\label{3-15}
\end{figure}

Next, we  fix either the $v_\uparrow$ or $v_\downarrow$
parameter while adjusting the other to incrementally vary the difference
$|v_\downarrow-v_\uparrow|$. In Fig.  \ref{3-16}(a),    with $N=15$,   $V_1=3w_\uparrow$,   $v_\uparrow=0. 1w_\uparrow$,   $w_\uparrow=w_\downarrow$,    we uniformly increase $v_\downarrow$ from $0. 2w_\uparrow$ to $0. 5 w_\uparrow$,    thereby increasing the difference $|v_\downarrow-v_\uparrow|$. Using the same method as before,    we obtain oscillation trajectory of the $\langle x \rangle$  until $t=5\times10^5$ and transform it from the time domain to the frequency domain via Fourier transform to extract the main oscillation frequency components.  The obtained oscillation frequency is then fitted with the corresponding $|v_\downarrow-v_\uparrow|$ revealing a linear  decay of oscillation frequency as a function of $|v_\downarrow-v_\uparrow|$, where slope value is -367.86 and intercept value is 141.25 .

In Fig.  \ref{3-16} (b),    except for the fixed $v_ \downarrow=0. 5w_ \downarrow $,    all other parameters are the same as in Fig 14(a).  We uniformly increase $v_ \uparrow $ from $0. 2w_ \uparrow $ to $0. 5w_ \uparrow $,    thereby the difference between $| v_ \downarrow - v_ \uparrow | $ gradually decreases to 0.  We obtained the oscillation frequency using the same method and fitted it to the corresponding $|v_\downarrow-v_\uparrow|$.  The linear  decay  is the same,    except that the slope value -821.43 and intercept value 293.93 of the fitting relation are different from the Fig 14(a).

Unlike varying $|v_\downarrow-v_\uparrow|$, adjusting $|w_\uparrow-w_\downarrow|$ significantly affects the oscillatory motion of particles,    as demonstrated in Fig. \ref{3-15}.  We compared two distinct  $|w_\uparrow-w_\downarrow|=0. 1$ and $|w_\uparrow-w_\downarrow|=0. 05$. Within the same time scale,    particles oscillate slowly across the entire chain when
$|w_\uparrow-w_\downarrow|=0. 1$, whereas they oscillate only within a single site at each end when $|w_\uparrow-w_\downarrow|=0. 05$ showing strong localization. 
The primary cause is the decrease in $|w_\uparrow-w_\downarrow|$, which leads to a reduction in the decay length and consequently a slower oscillation rate.

\section{Conclusion}

In conclusion, our study focused on a two-body model  within the double-chain SSH Hubbard model for a quantum dot system. This model allows for the individual adjustment of intracell and intercell tunneling parameters of distinct spin particle. Through exact diagonalization of Hamiltonian  under given parameters, we derived the real-space energy spectrum. We further conducted energy and wave function analyses on various states, including non-interacting bulk states, non-interacting bulk-edge states, non-interacting edge-edge states, correlated bulk states, and correlated edge states that appeared in the energy spectrum.

We concentrated mainly on examining the properties of energy, decay length, and dynamical behavior of the correlated edge states $d_2$ under various parameters, including interaction strength $V_1$ and tunnelings $v_\uparrow$, $v_\downarrow$, $w_\uparrow$, $w_\downarrow$. We found that different spin tunneling values can lift the energy degeneracy and lead to energy shifts in the correlated edge states. The influence of interaction on the decay length of correlated edge states is characterized by the observation that stronger interactions result in longer decay lengths of the states. The decay length of the state decreases as the difference in intracell $|v_\downarrow-v_\uparrow|$ or the difference in intercell tunneling  $|w_\uparrow-w_\downarrow|$ increases. The decay length is strongly correlated with the dynamical behavior of two particles. As the decay length increases, the frequency of the oscillation decreases. This result has been confirmed through the observation of the expectation coordinate values of the particles. This work has potential implication for the experimentally investigating electron dynamics and topological properties in a double array of quantum dots and for many-body physics in quantum dot systems.

\appendix
\section{Two-body spectrum in momentum space}
\begin{figure}[htpb]
	\centering
\includegraphics[width=9.00cm,   height=8cm]{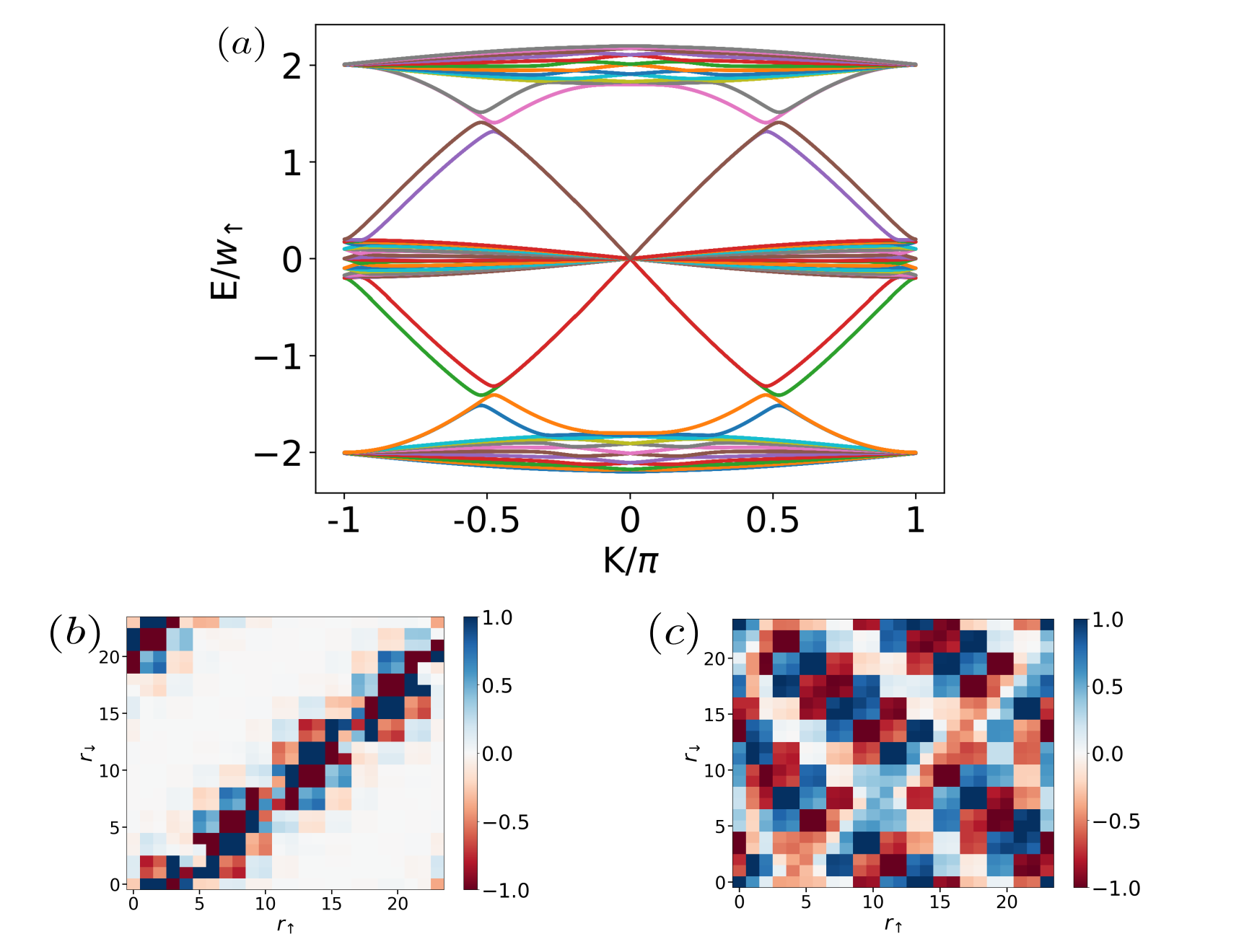}
	\captionsetup{justification=raggedright,    singlelinecheck=false}
	\caption{The energy spectrum and partial wavefunction distributions of two-particle states as a function of the center-of-mass momentum \(K\)  for PBC,    with no interactions \(V_1  = 0\).   (a) \(N = 12\), $v_\uparrow = v_\downarrow = 0.1w_\uparrow$, $w_\uparrow = w_\downarrow$.  The broad energy bands correspond to the non-interacting  bulk states,    while the isolated narrow bands correspond to the non-interacting  bound states.  (b) and (c) For \(K = 1\),    the wavefunction distributions are shown for the state at \(E \sim 1. 99w_\uparrow\) and at \(E \sim 1. 63w_\uparrow\) respectively.  The coordinate axis represent the lattice site positions of the two electrons.  The color indicates the wavefunction probability amplitude. For improved color prominence,    the data values were magnified by L times when generating the plot. }
	\label{3-2}
\end{figure}
When the parameters in the Eqs. (\ref{3. 1. 3})  are fixed,    the Hamiltonian can be exactly diagonalized to obtain the energy spectrum as a fuction of $K$.

Two-body spectrum for PBC without any interaction and the wavefunctions of the two-particle states in spectrum are illustrated in Fig. \ref{3-2}.  In the figure (a),    the broad bands near energies \(\pm2w_\uparrow\) and 0 correspond to the non-interacting two-particle bulk states,    where the wavefunction distribution indicates that the two particles are free particles and randomly occupy all the lattice sites along the chain.  The isolated narrow bands within the broad bands gap are composed of bound states,    where the wavefunction of the two particles are primarily localized on neighboring lattice sites.  These bound states are not caused by interactions but rather by the periodic boundary conditions. 

\section{Energy spectrum and wave function with identical spin tunneling}

\begin{figure}[htpb]
	\centering
    \includegraphics[width=9.00cm,   height=5cm]{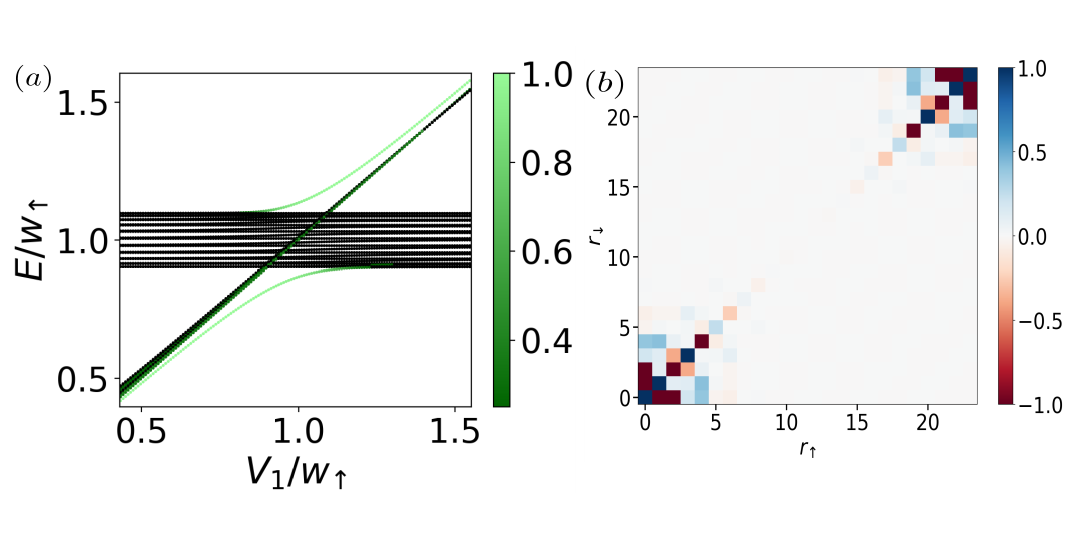}
	\captionsetup{justification=raggedright,    singlelinecheck=false}
	\caption{Energy spectrum under OBC and wave function with identical spin tunneling, where $N=12$,  $v_\uparrow=v_\downarrow=0.1 w_\uparrow$, $w_\uparrow$=$w_\downarrow$ (a) The energy spectrum of the $d_2$ band near $V_1=w_\uparrow$.  There is a doubly degenerate strongly localized edge state (symmetrically inclined  green curves). (b) The correlated edge state at  energy  $E\sim0.57w_\uparrow$. }
	\label{17}
\end{figure}

We focus on the energy spectrum region near $V_1=w_\uparrow$
for the $d_2$ state under OBC in the case of identical spin tunneling $v_\uparrow=v_\downarrow$,   $w_\uparrow=w_\downarrow$, as depicted in Fig. \ref{17}. The purpose is to compare with the results of different spin tunnelings.

\section{The projections along the diagonal of the wave functions for different $V_1$ values}
We show the projections along the diagonal $x=y$ of pictures (a) to (d) in Fig. \ref{fig:3-11}. The red and blue color bars indicate the absolute values of the probability amplitudes for the two correlated particles occupying even and odd lattice sites,   respectively. 

\begin{figure}[htbp]
	\centering
	\includegraphics[width=0.9\linewidth]{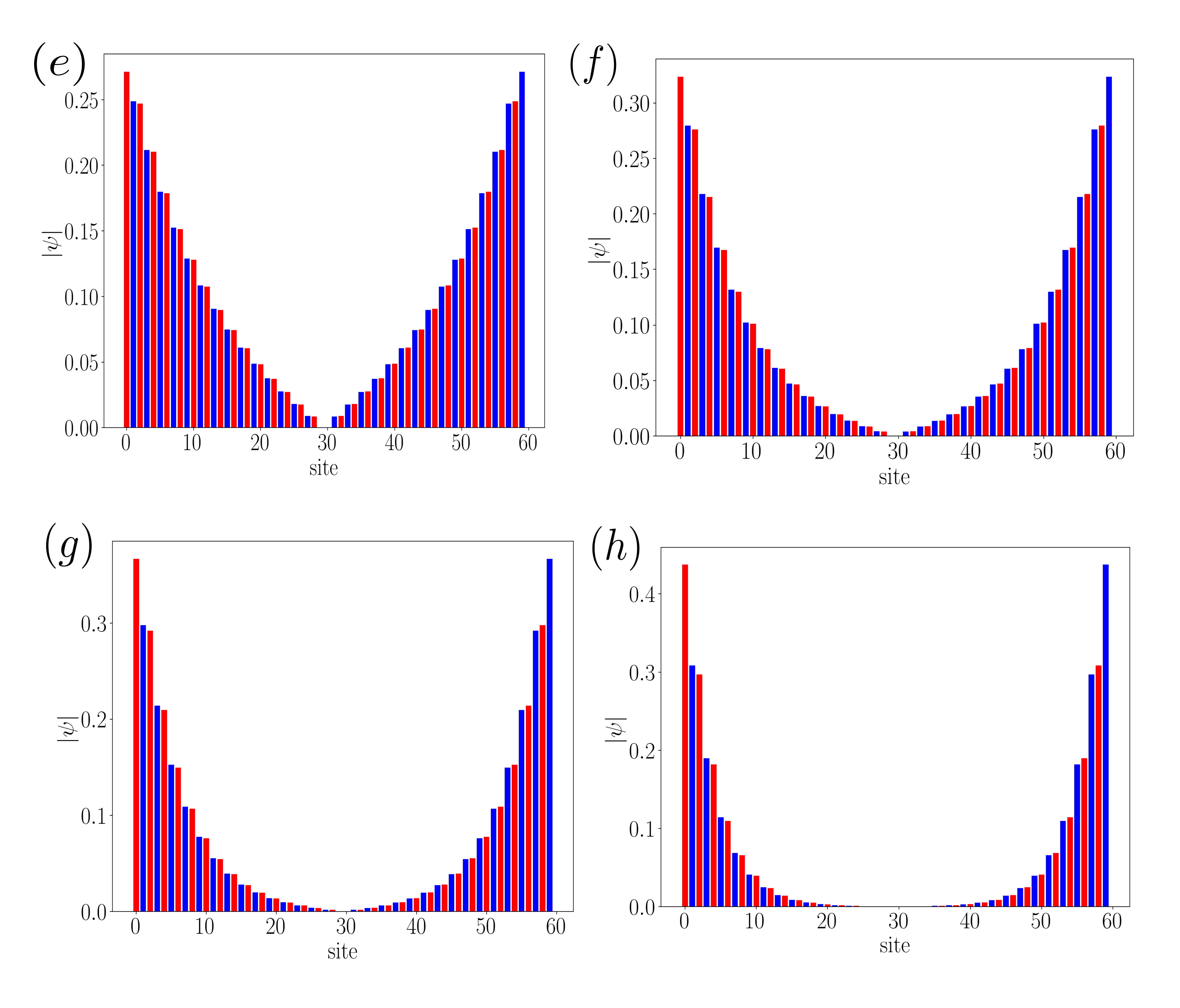}
	\captionsetup{justification=raggedright,    singlelinecheck=false}
	\caption{The wavefunctions projections along the diagonal of the correlated two-particle edge states at different $V_1$ values.  $N=30$,  $v_\uparrow=0. 1w_\uparrow$,    $v_\downarrow=0. 3w_\downarrow$,    $w_\uparrow=w_\downarrow$. 
		(e) to (h) show the projections along the diagonal $x=y$ of pictures (a) to (d) in Fig. \ref{fig:3-11},    intuitively illustrating the feature of the edge state decaying from the end lattice sites. 
		The $x$-axis denotes the positions of  lattice sites on the chain,    and the $y$-axis denotes the absolute values of the probability amplitudes for particles occupying a specific lattice site.  Red and blue bars indicate the two correlated particles occupying even and odd lattice sites,    respectively. 
		When $V_1=2. 5w_\uparrow$,    the decay length exhibits an exponential decay,    whereas when $V_1=3. 5w_\uparrow$,    the decay length increases and does not follow an exponential decay. }
	\label{18}
\end{figure}

\acknowledgements
This work is supported by the National Natural Science
Foundation of China (Grants No. 11904157, No.
62174076, and No. 92165210), Shenzhen Science and Technology
Program (Grant No. KQTD20200820113010023),
and Guangdong Provincial Key Laboratory (Grant No.
2019B121203002).

\nocite{*}

\bibliography{apssamp}

\end{document}